\documentclass[aip,jcp,
amsmath,amssymb,
floatfix,
reprint,
]{revtex4-1}

\usepackage[utf8]{inputenc} 
\usepackage[T1]{fontenc} 
\usepackage{verbatim}

\usepackage[table]{xcolor} 
\usepackage{graphicx}

\usepackage{dcolumn}
\usepackage{bm}

\usepackage{listings}

\usepackage[version=3]{mhchem}
\usepackage{units}
\usepackage{siunitx}

\usepackage{xspace}
\usepackage{todonotes}

\graphicspath{{pics/}}

\renewcommand{\vec}{\boldsymbol} 

\newcommand{\eq}{Eq.\xspace}
\newcommand{\Eq}{Eq.\xspace}
\newcommand{\fig}{Fig.\xspace}
\newcommand{\tab}{Table\xspace}

\newcommand{\Fig}{Fig.\xspace}
\newcommand{\ie}{\mbox{i.\,e.\ }}
\newcommand{\eg}{\mbox{e.\,g.\ }}

\newcommand{\Ref}{Ref.\xspace}
\newcommand{\lit}[1]{\Ref\citenum{#1}\xspace}

\newcommand{\SOP}{{SoP}\xspace}

\newcommand{\FGH}{{FGH}\xspace}

\newcommand{\pvb}{PvB\xspace}
\newcommand{\pW}{pW\xspace}

\newcommand{\nMean}{\overline}

\newcommand{\matrgreek}[1]{\ensuremath{\pmb{#1}}}

\newcommand{\conj}[1]{\ensuremath{{#1}^\ast}}

\newcommand{\matr}[1]{\textbf{#1}}

\newcommand{\ket}[1]{\ensuremath{\left| {#1} \right \rangle }}

\newcommand{\braket}[2]{\ensuremath{\left \langle #1 \vphantom{#1 #2}
    \left | \, #2 \vphantom{#1 #2} \right . \right \rangle }}

\newcommand{\matrixe}[3]{\ensuremath{ \left \langle{#1} \left| \vphantom
        {#1 #3} {#2} 
\right| {#3} \right \rangle }}

\newcommand{\sbraket}[2]{\ensuremath{ \langle #1 | \, #2  \rangle }}

\newcommand{\sket}[1]{\ensuremath{  | {#1} \rangle}}
\newcommand{\smatrixe}[3]{\ensuremath{ \langle{#1} | \vphantom
        {#1 #3} {#2} 
| {#3} \rangle }}

\newcommand{\partd}[2]{\ensuremath{ \frac{\partial {#1}}
{\partial {#2}} }}

\newcommand{\partdd}[2]{\ensuremath{ \frac{\partial^2 {#1}}
{\partial {#2}^2} }}

\newcommand{\ii}{\ensuremath{\mathrm{i}}}

\newcommand{\dd}{\ensuremath{\mathrm{d}}}

\definecolor{CBred}{RGB}{215,25,28}

\usepackage{tensor}
\begin{document}

\title{Efficient molecular quantum dynamics in coordinate and phase space using pruned bases}

\author{H.~R.~Larsson}
\email{larsson@pctc.uni-kiel.de}
\author{B.~Hartke}
\affiliation{Institut für Physikalische Chemie, Christian-Albrechts-Universität zu Kiel, 24098 Kiel, Germany}
\author{D.~J.~Tannor}
\affiliation{Department of Chemical Physics, Weizmann Institute of Science, 76100 Rehovot, Israel}
\date{\today}

\begin{abstract}
We present an efficient implementation of dynamically pruned quantum dynamics,
both in coordinate space and in phase space. We combine the ideas behind the biorthogonal von Neumann basis (\pvb) with the orthogonalized momentum-symmetrized Gaussians (Weylets) to create a new basis, projected Weylets, that takes the best from both methods. We benchmark pruned {time-dependent} dynamics using phase-space-localized \pvb, projected Weylets, and coordinate-space-localized DVR bases, with real-world examples in up to six dimensions. {For the examples studied,} coordinate-space localization is {the} most important {factor} for efficient pruning {and the} pruned dynamics is much faster {than the} unpruned, exact dynamics. Phase-space localization is useful for more demanding dynamics where many basis functions are required. There, projected Weylets offer a more compact representation than pruned DVR bases.
\end{abstract}

\maketitle

\section{Introduction}

Chemical reaction dynamics can be studied theoretically by molecular quantum dynamics.\cite{tannor_book} Especially, quantum effects like resonances and tunneling can be very important for the correct description of molecular processes in chemical reactions.\cite{H2_F_low_temp_sims_2014,H_CH4_manthe_2004,F_H2O_otto_2014,CH4F_manthe_2013}
Classical or semi-classical methods may fail to describe this.
However, the exact treatment of molecular quantum dynamics is hampered by the
exponential scaling of the underlying direct-product basis with the
dimensionality of the system. Exact quantum dynamics based on a
direct-product basis is now possible for up to five-atomic systems  
or nine degrees of freedom.\cite{C2H_H2_zhang_2016,H2_NH2_guo_2014}
Dynamics in reduced dimensionality is possible for six atoms and has been successfully
applied.\cite{six_atom_dyn_reduced_dim_clary_2000,H_CD4_reduced_dim_zhang_2007,red_dyn_h_transfer_descouter-lecomte_2002}  
However, in reduced dimensionality dynamics, the explicitly treated degrees of
freedom and the methods for approximate treatment of the other degrees of
freedom have to be selected and tested carefully, otherwise the results may 
deviate qualitatively from full-dimensional computations.\cite{CH4_H_full_versus_reduced_dim_qdyn_manthe_2001,CH4_H_reduced_dim_vs_full_dim_classical_lendvay_2016}

A physically motivated route towards reducing or even overcoming the exponential scaling aims
at representing the wavefunction not everywhere in configuration space but only
where it is needed. In a very general and hence very robust sense, this is
certain to incur huge savings, since in typical chemical situations the
vibrational wavefunction has many degrees of freedom but is narrowly
confined in most of them -- simply because chemistry is not about
many electrons and bare nuclei colliding at high speeds (which corresponds to
the wavefunction covering large parts of coordinate space and/or phase space)
but rather about stable molecules undergoing well-defined reactions, which
means that chemical energies are just above one or a few barriers. There
simply is not enough energy to break arbitrary bonds in a reactant molecule.
Hence, the mathematically simple and thus appealing direct-product bases already are wasteful
for dynamics of small molecules and become even more wasteful for larger
molecules. 

A conceptually elegant way to exploit these characteristics is to employ a
basis representation of the wave packet in which the basis functions
themselves move around and follow the wave packet, \eg time-dependent Gaussians.
These Gaussians are either moved in time by classical mechanics or by proper quantum treatment. There is a plethora of established methods like G-MCTDH,\cite{G-MCTDH_Burghardt_1999,G-MCTDH_layer_Burghardt_2013} its cousin variational Multiconfigurational Gaussians, vMCG,\cite{vMCG_Burghardt_2004,vMCG_rev_lasorne_2015} the matching-pursuit algorithm, MP-SOFT,\cite{matching_pursuit_SPO_batista_2003}
multiple spawning dynamics\cite{aims_martinez_1996,aims_validation_martinez_1998} or the coupled coherent state approach.\cite{coupled_coherent_states_rev_shalashilin_2004}
However, these methods are either not fully exact or often suffer from numerical difficulties.\cite{vMCG_rev_lasorne_2015}

Another possibility is to use an optimal time-dependent direct-product basis
built upon a larger time-independent primitive basis. The outcome is then the multi-configurational time-dependent Hartree (MCTDH) algorithm. \cite{mctdh_cederbaum_1990,mctdh_rev_meyer_2000}
MCTDH and especially its multilayer variant
(ML-MCTDH)\cite{ml_mctdh_thoss_2003,ml_mctdh_manthe_2008,ml_mctdh_meyer_2011}
is often an efficient way to reduce the exponential scaling. Although a
direct-product basis is still used, the number of needed basis functions is
drastically reduced. This achievement comes with a much more complicated algorithm and it may work less well for dynamics with strongly coupled modes. Further, to find proper mode combinations in normal MCTDH (to optimize multidimensional bases for efficiency) or tree structures in ML-MCTDH is a nontrivial task.\cite{ml_mctdh_H_CH4_manthe_2012,mlmctdh_versus_mctdh_H2COO_meyer_2014}

Using lower-dimensional direct-product bases to create contracted basis functions has been found to be very successful for quantum (ro-)vibrational computations.\cite{ray_dvr_light_1986,ray_dvr_appl_light_1987,vib_spectra_contraction_gazdy_1990,basis_contraction_rovib_handy_1992,contracted_basis_vibrational_spectrum_HFCO_leforestier_2000,contraction_vibrational_generalized_coordinate_dvr_luckhaus_2000,contracted_bases_lanczos_carrington_2002,asaf_thesis}
A contracted basis was even used for computing vibrational energy levels of 12-dimensional \ce{CH5+}.\cite{CH5+_vib_carrington_2008}

Yet another possibility, which certainly can be combined with the MCTDH \emph{ansatz}, is to prune a time-independent direct-product basis and choose only those functions that are necessary to describe the wave function.
This approach is very successful for solving the time-independent Schrödinger equation, \ie retrieving the eigenfunctions. It can be simply implemented by discarding functions that are located at points with high potential,\cite{ray_dvr_light_1986,ray_dvr_appl_light_1987}
or by more advanced techniques like simultaneous diagonalization,\cite{phase_space_localised_DVR_carrington_2005,cubature_tannor_2005,multidimensional_dvr_cubature_tannor_2006} utilization of phase-space-structured basis functions\cite{eigenstates_fluoroform_efficient_vib_basis_lung_1997,vibrational_selected_ci_vmwci_lievin_2007,benzene_hybrid_truncation_scheme_HO_basis_poirier_2015,pvb_LiCN_tannor_2014} 
or sparse grids.\cite{multidim_dvr_simultaneous_diagonalisation_carrington_2004,nonprod_grid_carrington_2009,sparse_grid_smolyak_nauts_2014}
It has also been used for time-independent scattering simulations.\cite{sinc_dvr_miller_1992}

For solving the time-dependent Schrödinger equation, \eg for simulating
time-dependent quantities in a chemical reaction, 
pruning is much more complicated because the wavepacket moves in time. One may use static pruning.\cite{pruning_tdse_carrington_2010,sparse_grid_fft_gradinaru_2007}
Especially, an L-shaped selection of basis functions localized in coordinate
space has been used very successfully in quantum scattering simulations.\cite{H2_OH_diatom_diatom_reaction_treatment_zhang_1994}
However, a more general pruning scheme that really adapts the basis in time should be much more efficient. 
This has been implemented and thoroughly tested by Hartke and coworkers for coordinate-space localized Gaussians\cite{proDG_hartke_2006} and orthogonalized Gaussians based on collocation.\cite{proDG_hartke_2008}
McCormack showed that it is a possible avenue for coordinate-space-localized
discrete variable representation (DVR) functions and phase-space-localized
functions but did not present an implementation.\cite{pruning_mccormack_2006}
Pettey and Wyatt implemented pruning by a moving
grid.\cite{pruning_wyatt_2006,pruning_wyatt_2007} An additional advantage of
dynamical pruning is the ability to compute the needed points of the
potential energy surface (PES) on the fly during the dynamics, instead of
pre-computing the PES prior to the dynamics, which makes it hard to avoid
computing PES point that are never needed and to avoid exponential scaling
already at this stage.

Because the wave packet is most often not only confined in coordinate space
but also exhibits structured motion in phase space (combined position $x$ and
momentum $p$), pruning based on phase-space-localized functions 
generally gives more compact representations. However, obtaining an accurate and
prunable basis of a well-defined rectangular grid of phase-space-localized
functions is far from trivial\cite{semicl_gauss_heller_1979} and has only been 
established in the last few
years.\cite{weylet_1_poirier_2003,weylet_2_poirier_2004,weylet_3_poirier_2004,pvb_tannor_2012}
Until now, the so-called Weylets and momentum-symmetrized Gaussians by Poirier \emph{et al.} have not been used for quantum dynamics simulations but only for computing eigenstates. There, they have been shown to be very successful.\cite{weylet_3_poirier_2004,weylets_Ne2_poirier_2006,symmetr_gauss_weylet_poirier_2012,symmetr_gauss_appl_CH2NH_poirier_2014,symmetrized_gaussians_acetonitrile_poirier_2015,benzene_hybrid_truncation_scheme_HO_basis_poirier_2015}
The periodic von Neumann basis with biorthogonal exchange, \pvb, from Tannor and coworkers has been both successfully applied to vibrational eigenstates\cite{pvb_LiCN_tannor_2014} and to electronic dynamics.\cite{pvb_edyn_takemoto_tannor_2012,pvb_edyn_tannor_2015}
It has not yet been used in {time-dependent} molecular quantum dynamics.

In this contribution, we develop a very efficient implementation of dynamical,
time-dependent pruning. We benchmark several bases using two-, three- and six-dimensional examples. The bases we employ are coordinate-space-localized DVR functions, the biorthogonal von Neumann basis, \pvb, and a novel method, called projected Weylets, which combines the idea behind \pvb with the Weylet basis to create a basis that inherits the advantages of the \pvb methods without inheriting its disadvantage, namely nonorthogonality.

Our implementation is based on Hamiltonians that are a sum of products of one-dimensional operators (\SOP).
{Poirier and coworkers used \SOP Hamiltonians for their computations using Weylets and symmetrized Gaussians.\cite{weylet_3_poirier_2004,symmetr_gauss_weylet_poirier_2012,symmetr_gauss_appl_CH2NH_poirier_2014,symmetrized_gaussians_acetonitrile_poirier_2015,benzene_hybrid_truncation_scheme_HO_basis_poirier_2015}}
{Moreover}, Brown and Carrington have used the mo\-men\-tum-symme\-tri\-zed Gaussians of Poirier together with \SOP Hamiltonians for vib\-ra\-tio\-nal com\-pu\-tations.\cite{symmetrized_gaussians_sop_carrington_2016} 
Many {other} methods require \SOP Hamiltonians or are much more efficient if \SOP Hamiltonians are utilized, \cite{mctdh_rev_meyer_2000,vcc_rev_christiansen_2012,cp_decomposition_carrington_2014}
and there are  many algorithms available to fit the Hamiltonian in a \SOP form efficiently, notably potfit and variants thereof,\cite{potfit_meyer_1996,multigrid_potfit_meyer_2013,mctdh_multilayer_potfit_otto_2014} neural networks,\cite{pes_neural_network_sum_of_products_carrington_2006,sop_pes_neural_networks_zhang_2014} or some interpolation techniques.\cite{pes_sum_of_products_smolyak_carrington_2015,pes_sop_from_multimode_fit_rauhut_2016}
Note, however, that pruned DVR bases do not require \SOP Hamiltonians (to be precise: A \SOP form of the potential) and we show that these bases can be very successfully pruned.

The remainder of this article is organized as follows: Section \ref{sec:theory_overview} gives an overview of exact quantum dynamics with a direct-product basis. A review of the \pvb method, its drawbacks in high dimensions, and the Weylets follow in section \ref{sec:pvb} and \ref{sec:weylets}, respectively. We present our new projected Weylets in section \ref{sec:projected_weylets}, followed by details of the numerical implementation in section \ref{ch:pW_numImpl} and the appendix. The mentioned methods and a pruned DVR basis are tested in section \ref{sec:appl}. Section \ref{sec:conclusion} summarizes the most important points of this work and gives an outlook.

\section{Theory}
\subsection{Overview}
\label{sec:theory_overview}
The standard approach in molecular quantum dynamics for solving the time-dependent Schrödinger equation is to expand the $D$-dimensional wave function in a direct product of time-independent basis functions  $\left\{\ket{\chi_{i_\kappa}^{(\kappa)}}\right\}_{i_\kappa=1}^{n_\kappa}$,\cite{mctdh_rev_meyer_2000,tannor_book} 
\begin{equation}
 \ket{\Psi(t)} =  \sum_{i_1=1}^{n_1} \dots \sum_{i_D=1}^{n_D} a_{i_1i_2\dots i_D} \bigotimes_{\kappa=1}^D \ket{\chi_{i_\kappa}^{(\kappa)}}.\label{eq:wavepacket_form}
\end{equation}
The size of the coefficient tensor, $\vec a$, scales as $\prod_{\kappa=1}^D n_\kappa = \nMean{n}^D$, %
where $\nMean n$ is the geometric mean of the number of basis functions.
A discrete variable representation (DVR) is often used as the underlying basis, $\left\{\sket{\chi_{i_\kappa}^{(\kappa)}}\right\}$.\cite{tannor_book}
This \emph{ansatz} is then inserted into the time-dependent Schrödinger equation,
\begin{equation}
 \ii \partd{}{t} \ket{\Psi(t)} = \hat H \ket{\Psi(t)},
\end{equation}
to obtain
\begin{align}
 \ii \partd{}{t} \vec a(t) &= \matr H \vec a(t), \label{eq:tdse_a}\\
 H_{IJ} &= \bigotimes_{\kappa=1}^D\bigotimes_{\tau=1}^D \matrixe{\chi_{i_\kappa}^{(\kappa)}}{\hat H}{\chi_{j_\tau}^{(\tau)}},
\end{align}
where we have introduced multiindices $I\equiv i_1i_2\dots i_D$. Throughout the text, we use atomic units unless stated otherwise. 
\Eq\eqref{eq:tdse_a} can be solved using standard integrators that require
many tensor transformations of the type $\matr H \vec a$. 
Using multiindices, this tensor transformation can be considered as a matrix-vector product. The terms {``tensor transformation'' and ``matrix-vector product''} will be taken as a synonym throughout the text.

The direct-product basis is conceptually simple and easy to implement. A DVR basis allows further simplifications due to the diagonality of the potential operator $\hat V$ which leads to diagonal matrices{, i.\,e.,} $V_{IJ} \propto \delta_{IJ}$. 
The problem is the exponential scaling of the size of the coefficient tensor, $\vec a$. 
When using a DVR as the underlying basis, the basis functions are localized in coordinate space. Since the wavepacket is normally not spread over the whole multidimensional coordinate space, the coefficient tensor $\vec a$ is sparse and it is possible to prune the basis. 
By using a phase-space-localized basis, the coefficient tensor gets sparser and pruning is often more efficient. The next sections will mostly deal with phase-space-localized basis functions.
\subsection{Biorthogonal projected von Neumann basis}
\label{sec:pvb}
\subsubsection{Theoretical foundations}
A grid of phase-space-localized basis functions
is established with von Neumann functions\cite{neumann_1931,neumann_book}
\begin{align}
\!\!\!\braket{x}{\tilde g_{nl}}\! &=\! \left(\frac{2\alpha}{\pi}\right)^{\frac14}\!%
   \exp\left[-\alpha(x-x_n)^2\!\!+\! \ii\! \times\! p_l  (x - x_n)\right],\\
\alpha &= \frac{\Delta p}{2\Delta x}.
\end{align}
The basis functions are localized at $(x_n,p_l)$ in phase space with widths $\Delta p$ and $\Delta x$ in $p$ and $x$. They are placed on a rectangular lattice in phase space with rectangles of widths $\Delta p$ and $\Delta x$ such that $\Delta x \Delta p = 2\pi$, which is the size of a unit cell in phase space.
This assures that the basis is complete but not overcomplete.
The von Neumann basis was used by Davis and Heller but they found very poor convergence.\cite{semicl_gauss_heller_1979}
{This may be understood as a consequence of the theorem of Balian and Low which states that a phase-space-localized basis is incompatible with completeness for all practical purposes.\cite{balian_low_theorem_balian_1981,balian_low_theorem_low_1985,balian_low_theorem_proof_battle_1988}}
Shimshovitz and Tannor have shown that \emph{projecting} the von Neumann functions to a different (DVR-like) basis $\{\ket{\chi_i}\}$ solves this 
problem:\cite{pvb_tannor_2012}
\begin{equation}
  \ket{g_i} = \sum_{j}\ket{\chi_j} \braket{\chi_j}{\tilde g_i} = \sum_j %
\ket{\chi_j} \sqrt{\omega_j}\braket{x_j}{\tilde g_i} \label{eq:pvN_def}
\end{equation}
Here, the von Neumann functions are labeled by a multiindex.
The last equality comes from the DVR properties of $\{\ket{\chi_i}\}$.
$\omega_j = W_j/\omega(x_j)$, where $W_j$ is the quadrature weight of the DVR point $x_j$ and $\omega(x_j)$ the weight function of the underlying DVR polynomials.\cite{tannor_book}
Since \eq\eqref{eq:pvN_def} denotes just a similarity transformation, utilization of $\ket{g_i}$ to solve the Schrödinger equations gives exactly the same eigenvalues as with the DVR basis.
However, the basis $\{\ket{\chi_i}\}$ has to occupy the same area in phase space to render the transformation bijective. The Fourier Grid Hamiltonian (FGH) DVR fulfills this property and is thus often used.\cite{fgh_marston_balint-kurti_1989,tannor_book} This makes the $\ket{g_i}$ periodic. However, other (nonperiodic) DVR bases like the sinc DVR or Gauss-Legendre DVR (for angular coordinates) are possible (see also section \ref{sec:NO2}).\cite{sinc_dvr_miller_1992,tannor_book,pvb_LiCN_tannor_2014} Other non-DVR bases or even simple Newton-Cotes quadrature work as well.

These basis functions are localized in phase space, but their coefficients are not, because they include a linear combination of the inverse of the overlaps of the $\ket{g_j}$:
\begin{align}
\!\!\ket{\Psi} &= \sum_{m}  \ket{g_m}{} ^ga_m\! = \sum_{m=1}^N \ket{g_m} \sum_{n=1}^N [^{g}S^{-1}]_{mn}\!
\braket{g_n}{\Psi},\label{eq:psi_exp_pvN}\\
^{g}\matr S &= \matr G^\dagger \matr G.
\end{align}
Left superscripts indicate the employed basis.
Because $^g\matr S^{-1}$ is not sparse, the coefficients $^ga_m$ are dense and pruning is not possible.
Shimshovitz and Tannor solved this problem by transforming to the biorthogonal basis where  
\begin{align}
  \ket{b_n} &= \sum_{m=1}^N \ket{{g}_m}[^gS^{-1}]_{mn},\label{eq:pvb_def} \quad \braket{b_n}{g_m} = \delta_{mn},\\
  \ket{\Psi} &= \sum_{m=1}^N \ket{b_m}{} ^ba_m = \sum_{m=1}^N  \ket{b_m}\braket{{g}_m}{\Psi}.\label{eq:psi_exp_pvb}
\end{align}
The coefficients in this representations are just the overlap of the wavefunction with phase-space-localized $\ket{g_i}$ and hence sparse. Note that $\braket{x}{b_i}$ is anything but localized{, which is a manifestation of the Balian-Low theorem.\cite{balian_low_theorem_proof_battle_1988,balian_low_generalisation_janssen_1993}}
Indeed, it does not matter whether the basis is localized but whether the coefficients are. $\braket{b_i}{\Psi}$ is actually the expansion coefficient in the $\ket{g}$ representation, compare with Eqn.~\eqref{eq:psi_exp_pvN} and \eqref{eq:pvb_def}.
The basis $\{\ket{b_i}\}$ is called the periodic von Neumann basis with
biorthogonal exchange (\pvb). Because the basis was initially based on FGH
functions, it was termed periodic. However, since also other basis functions
can be used and because the essential part in \eq\eqref{eq:pvN_def} is the
projection onto the DVR basis and not its periodicity, the basis may also be dubbed \emph{projected} von Neumann basis.

Because the basis is nonorthogonal the Schrödinger equation takes the form of
\begin{equation}
  \ii ^b\matr S \partd{}{t}{}\, ^b\vec{a} = {}^b\matr H\,{} {^b\vec{a}}{},
\end{equation}
where $^b\matr H$ is just the {congruence}-transformed DVR Hamiltonian ${}^\chi\matr H$:\cite{pvb_math_tannor_2016}
\begin{equation}
 ^b\matr H = \matr B^\dagger {}^\chi{}\matr H \matr B,\quad B_{ji} =\sqrt{\omega_j} \braket{x_j}{b_i}
 \label{eq:pvb_H_trafo}
\end{equation}

\subsubsection{Drawbacks in high dimensions}
\label{ch:pvb_drawbacks}
In the following, we will assume that the Hamiltonian can be decomposed as a sum of direct products (\SOP) of one-dimensional matrices, 
\begin{equation}
{}^\chi{}\matr H = \sum_{l=1}^g \bigotimes_{\kappa=1}^D {}^\chi\matr h^{(\kappa,l)}.
\end{equation}
Each matrix $^\chi{}\matr h^{(\kappa,l)}$ needs then to be transformed as shown in \eq\eqref{eq:pvb_H_trafo}, which comes with essentially no computational cost. 
If the Hamiltonian possesses the \SOP form, the matrix-vector product scales as $\mathcal O(\nMean n^{D+1})$ instead of $\mathcal O(\nMean n^{2D})$, if the matrix-vector product or tensor transformation is done sequentially.\cite{DVR_matvec_calc_manthe_1990,DVR_matvec_calc_carrington_1993}
To give an example, consider a three-dimensional problem with $g=1$:
\begin{align}
  (\matr H \matr a)_{pqr} &= a'_{pqr} = \sum_{i=1}^{n_1}\sum_{j=1}^{n_2}\sum_{k=1}^{n_3}h^{(1)}_{pi}h^{(2)}_{qj}h^{(3)}_{rk} a_{ijk} %
  \nonumber
\\
&= \sum_{i=1}^{n_1}h^{(1)}_{pi}%
 \underbrace{\sum_{j=1}^{n_2} h^{(2)}_{qj} \underbrace{\sum_{k=1}^{n_3} h^{(3)}_{rk} a_{ijk}}_{\tilde{a}_{ijr}}}_{\tilde{\tilde{a}}_{iqr}}
 \label{eq:matvecprod_sequential}
\end{align}
The $\mathcal O(\nMean n^{D+1})$ scaling is achieved by first computing all needed values of $\tilde{a}_{ijr}$, reusing them for computing all needed values of $\tilde{\tilde{a}}_{iqr}$ and reusing those for the final transformation to $a'_{pqr}$.

If one \emph{prunes} the basis, some coefficients $a_{abc}$ are neglected and only $\widetilde{n}_\kappa < n_\kappa$ basis functions are used in dimension $\kappa$. The pruning can be formulated by introducing sets $\mathcal I$, $\mathcal J(a)$ and $\mathcal K(a,b)$ that store the indices of the used basis functions. \Eq\eqref{eq:matvecprod_sequential} then takes the form of
\begin{align}
  a'_{pqr} &= \sum_{i\in \mathcal I}h^{(1)}_{pi}\sum_{j\in \mathcal J(i)} h^{(2)}_{qj} \sum_{k\in \mathcal K(i,j)} h^{(3)}_{rk} a_{ijk} \\%
 &=  \sum_{i\in \mathcal I}h^{(1)}_{pi} \tilde{\tilde{a}}_{iqr},\quad \tilde{\tilde{a}}_{iqr} = \sum_{j\in \mathcal J(i)} h^{(2)}_{qj} \tilde{a}_{ijr},\quad \dots.
\end{align}
Wang and Carrington used a similar form for specific index ranges.\cite{efficient_matvec_pruned_carrington_2001} Here, no assumptions about the structure in the pruning are made.
In these expressions, values of $\tilde{\tilde{a}}_{iqr}$ are needed for the computation of $\matr{a}'$ that are not
included in the index sets, \ie $q$ may not be element of $\mathcal J(i)$ and $r$ may not be element of $\mathcal K(i,q)$. Then, in principle, almost all possible index combinations of $\tilde{\tilde{a}}_{iqr}$ need to be
evaluated. The same holds for the needed values of $\tilde{\matr a}$ for the computation of $\tilde{\tilde{\matr a}}$. However, if the coefficient
tensor is sparse and the Hamiltonian transforms it to the same sparse
representation, the
partially transformed tensors $\tilde{\vec a}$ and $\tilde{\tilde{\vec a}}$
are sparse as well and the values of coefficients which are not elements of the
employed set of the initial tensor can be neglected. The
favorable scaling of order $\mathcal O(\nMean{\widetilde{n}}^{D+1})$ results. An efficient implementation for this pruned tensor transformation is presented in the appendix.

Because the \pvb representation is nonorthogonal, matrix-vector products of type $\matr S^{-1} \matr H \vec a$ are needed. Here, we can \emph{not} assume that the partially transformed tensors $\tilde{\tilde{\vec a}}$ and $\tilde{\vec a}$ are sparse, because $\vec a' = \matr H \vec a$ transforms the basis into the non-sparse $\ket{g}$ representation [see text below \eq\eqref{eq:psi_exp_pvb}]:
\begin{equation}
 a'_I = \sum_{J} \smatrixe{b_I}{\hat H}{b_J} \braket{g_J}{\Psi} = \smatrixe{b_I}{\hat H}{\Psi}.
\end{equation}
Only the final multiplication with $\matr S^{-1}$ transforms the tensor back
into the sparse $\ket{b}$ representation. Hence, one has to either compute
\emph{all} needed partially transformed coefficients which at worst results in a $\mathcal
O\left(\nMean{n}^{D+1}\right)$ scaling instead of $\mathcal
O\left(\nMean{\widetilde{n}}^{D+1}\right)$, or one has to do the transformation in
\eq\eqref{eq:matvecprod_sequential} directly to obtain a $\mathcal
O\left(\nMean{\widetilde{n}}^{2D}\right)$ scaling. None of these options are
favorable. 

Further, the inverse of the pruned overlap matrix is not
decomposable into a \SOP form. In a full basis, the overlap matrix obeys the \SOP form and so does its inverse,
\begin{equation}
  \matr S =  \bigotimes_{\kappa=1}^D \matr S^{(\kappa)}.
\end{equation}
$\matr S$ is the multidimensional tensor and $\matr S^{(\kappa)}$ one-dimensional matrices. 
Pruning the basis means that some rows and columns are removed from $\matr S$. Then, the inverse {of $\matr S$} in the truncated basis has no structure to exploit. %
Either, one needs to compute it explicitly and store a huge matrix of
size $\nMean{\widetilde{n}}^{2D}$, or one needs to compute $\matr S^{-1} \vec a'$
\emph{via} direct algorithms like conjugate gradient that perform many $\matr
S \vec a'$ operations (typically more than $100$ for numerical accuracy). We
conclude that, due to its  phase space locality and DVR accuracy, the \pvb
representation has very appealing properties. However, its nonorthogonality
results in the unfavorable $\mathcal O\left(\nMean{\widetilde{n}}^{2D}\right)$ scaling when computing the matrix-vector product.

It is possible to approximate the inverse of the pruned overlap matrix by
pruning the inverse of the unpruned overlap matrix.\cite{pvb_math_tannor_2016,pvb_H2O_carrington_2015,symmetrized_gaussians_sop_carrington_2016}
Then, the matrix-vector product $\matr S^{-1} \matr H \vec a = \widetilde{\matr H} \vec a$ can be done in one step and the favorable $\mathcal O\left(\nMean{\widetilde n}^{D+1}\right)$ scaling is recovered.\cite{pvb_H2O_carrington_2015,symmetrized_gaussians_sop_carrington_2016}
The introduced approximation is, however, difficult to control and is only useful for {time-dependent} quantum dynamics if very low accuracy is passable.\cite{pvb_math_tannor_2016} It may, however, be useful for computing vibrational eigenstates where this approximation is not severe.\cite{pvb_math_tannor_2016,pvb_H2O_carrington_2015,symmetrized_gaussians_sop_carrington_2016}

\subsection{Weylet representation}
\label{sec:weylets}
In the preceding section, we have shown that the nonorthogonality causes
problems if the \pvb representation is employed for pruning. Due to the
Balian-Low theorem, an \emph{orthogonal} phase-space-localized basis is not
possible.\cite{balian_low_theorem_balian_1981,balian_low_theorem_low_1985,balian_low_theorem_proof_battle_1988} 
However, Wilson\cite{gabor_basis_wilson_1987} and Daubechies \emph{et al.}\cite{wilson_gabor_basis_journe_1991} have shown that it is possible to obtain a \emph{momentum-symmetrized} basis that can be orthogonalized without jeopardizing its locality, so called ``Weyl-Heisenberg'' wavelets (Weylets). Poirier refined this basis and found a simpler orthogonalization procedure.\cite{weylet_1_poirier_2003,weylet_2_poirier_2004,weylet_3_poirier_2004}
The momentum-symmetrized Gaussians are
\begin{equation}
\begin{split}
\sbraket{x}{\widetilde\phi_{nl}} =& \left(\frac{8\alpha}{\pi}\right)^{\frac14} \exp\left[-\alpha (x-x_n)^2\right]\times \\
&\sin\left[p_l \left( x - x_n  -  \sqrt{\frac{\pi}{8\alpha}}\right) \right].\label{eq:sG_def}
\end{split}
\end{equation}
They are localized in $x_n$ and $\pm p_l$ and real-valued.
The phase factor in the sine term is crucial to {greatly reduce linear dependencies}. To maintain completeness, the functions are further placed on a \emph{doubly dense} grid, \ie $\Delta x \Delta p= \pi$, and $p_l\neq 0$. For further details, we refer to \lit{weylet_2_poirier_2004,weylet_3_poirier_2004}.

Although the underlying basis is symmetrized in $p$, because it is {doubly dense} it can describe states that are not symmetric in $p$. 
{
One unit-cell on the upper (or lower) plane in phase space has contributions from \emph{two} momentum-symmetrized Gaussians and a proper linear combination 
can then describe states of arbitrary shape in phase space, as they occur in {time-dependent} quantum dynamics.
}

The symmetrized Gaussians are then orthogonalized \emph{via} symmetric Löwdin orthogonalization, \ie using $\matr S^{-1/2}$.\cite{lowdin_orthogonalization_lowdin_1950} Since this basis is infinite-dimensional, in principle, $\matr S^{-1/2}$ has to be infinite-dimensional as well. However, the basis actually sets up a ``tight'' frame, \ie the basis functions are as orthogonal as possible and the overlap between two functions decays exponentially.\cite{weylet_2_poirier_2004} It is thus sufficient to obtain $\matr S^{-1/2}$ for a reasonably large basis. Poirier and Salam have listed the needed values of $\matr S^{-1/2}$ to high accuracy in \lit{weylet_2_poirier_2004}. 

\begin{figure}
\includegraphics{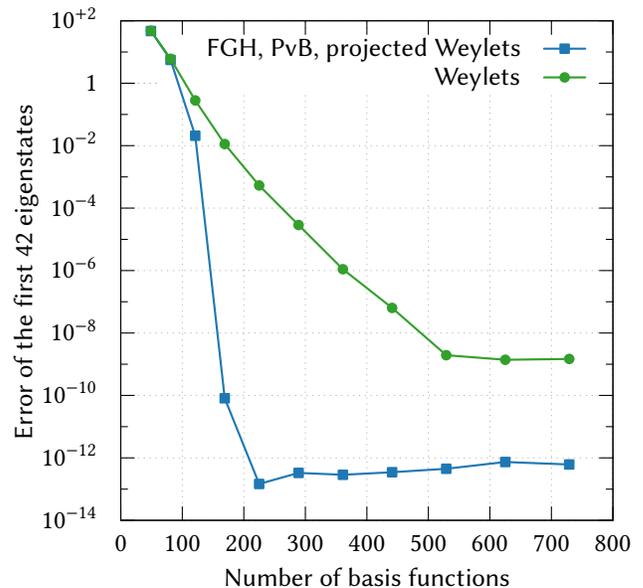}
\caption{Convergence of the {Weylets} (without {projection}) compared
  to a FGH DVR basis for the first $42$ states of the harmonic oscillator,
  $\hat H = (-\partial_x^2 + x^2)/2$. {The FGH-eigenvalues are identical to \pvb and projected Weylets (section \ref{sec:projected_weylets}).}
  {As the basis size is enlarged, both the $x$-range and the maximal momentum is increased such that a square area in phase space is covered. For each basis size, both bases span roughly the same phase-space area. 
  }}
 \label{fig:weylet_accuracy}
\end{figure}

{Weylets have the drawback that, compared to symmetrized Gaussians,} the transformation of the matrix elements to the Weylet basis can be quite cumbersome ({see section i of the supplementary information}), especially if the Hamiltonian does not obey a \SOP form.\cite{weylet_3_poirier_2004,weylets_Ne2_poirier_2006}
Even without pruning, the basis {in certain situations is} not as accurate as a Fourier basis. 
{A comparison of the convergence of Weylets, \pvb, FGH and projected Weylets (\pW, see section \ref{sec:projected_weylets}) for the eigenvalues of the harmonic oscillator is shown in \fig \ref{fig:weylet_accuracy}.
The error is defined by the Euclidian $L_2$ distance between the exact and the numerical energies.}
{It decreases much faster for FGH than for the Weylets. The asymptotic error of the Weylets stems from the limited accuracy (12 digits) of $\matr S^{-1/2}$ which has been taken from \lit{weylet_2_poirier_2004}. In practical calculations, this does not matter.  
Since both \pvb and projected Weylets are a similarity transformed FGH basis, they give the same eigenvalues as the FGH basis (apart from numerical noise).
In section {ii} of the supplementary information, we show further comparisons of Weylets against FGH, \pvb and \pW. We show there that for a pruned basis (no rectangular phase-space area), the error of the Weylets is almost identical to that of \pW.}

Poirier and coworkers avoid the transformation to the Weylet basis by
working solely with the nonorthogonal symmetrized Gaussians. They soften the $\mathcal
O\left(\nMean{\widetilde{n}}^{2D}\right)$ scaling by massive
parallelization.\cite{symmetr_gauss_weylet_poirier_2012,symmetr_gauss_appl_CH2NH_poirier_2014}
Note that this scaling is a function of the number of \emph{pruned} direct-product basis functions, $\nMean{\widetilde{n}}$.

\subsection{Projected Weylets}
\label{sec:projected_weylets}
{The relatively elaborate transformation needed for the creation of the Weylets (compared to the creation of momentum-symmetrized Gaussians)} can be {reduced} by combining them with {elements of} the projected von Neumann basis. First, the momentum-symmetrized Gaussians are projected onto the DVR lattice: 
\begin{equation}
  \ket{\phi_i} = \sum_{j}\ket{\chi_j} \sbraket{\chi_j}{\widetilde \phi_i} = \sum_j %
\ket{\chi_j} \sqrt{\omega_j} \sbraket{x_j}{\widetilde \phi_i}, \label{eq:psG_def}
\end{equation}
compare with \eq\eqref{eq:pvN_def}.
The projected Weylets, $\braket{x}{w_i}$, are then defined on the DVR grid as 
\begin{align}
 W_{ji}  &= \braket{x_j}{w_i},\\
 \matr W &= \matrgreek{\Phi}\,{} ^\phi \matr S^{-1/2},\\
 \Phi_{ji}&= \sqrt{\omega_j} \sbraket{x_j}{\widetilde \phi_i}, \quad ^\phi \matr S = \matrgreek \Phi^\dagger \matrgreek \Phi.
\end{align}
In \lit{symmetrized_gaussians_sop_carrington_2016}, Brown and Carrington mention in passing the possibility of projecting the momentum-symmetrized Gaussians onto a DVR grid, without implementing it. They do not consider orthogonalization.
Instead of employing momentum-symmetrized Gaussians $\sket{\widetilde \phi}$, one could also use the continuous Weylets as the initial basis. However, the numerical difference between these options is negligible. Instead of using orthogonalized functions, it is possible to use the biorthogonal basis and approximate the pruned inverse overlap matrix (see last paragraph in section \ref{ch:pvb_drawbacks}). Due to the more banded structure of the overlap matrix on a doubly dense grid, this approximation would be less severe than for \pvb.

The matrix representation of the Hamiltonian is again obtained by a similarity transformation of the DVR Hamiltonian, see \eq\eqref{eq:pvb_H_trafo}, and DVR accuracy is maintained. 
$\matr W$ represents now an orthogonal, momentum-symmetrized localized basis in phase space that has the same convergence properties as the underlying DVR basis, provided that $\{\sket{\widetilde\phi_i}\}$ and $\{\ket{\chi_i}\}$ span the same area in phase space. The only drawback that remains is the transformation of the Hamiltonian if it does not possess a \SOP form. However, the sparsity of the transformed Hamiltonian, $\matr W^\dagger {}^\chi{}\matr H \matr W$, and the transformation matrix, $\matr W$, can be exploited. This cannot be done with \pvb, because both $\matr B^\dagger {}^\chi{}\matr H\matr B$ and $\matr B$ are dense. The transformation for Hamiltonians without \SOP form will be addressed in future publications.

{To summarize, the projected Weylets {reduce the effort of the more involved} transformation of the Weylets because they are set up from a finite set of symmetrized Gaussians and not defined on an infinite plane, like Weylets {(see section i of the supplementary information)}. They show no reduction of accuracy compared to Weylets (see {\Fig\ref{fig:weylet_accuracy} and} Figure i in the supplementary information) and the generation of the basis is as simple as the generation of the \pvb basis.}
{Nevertheless, \pW should not be regarded as a replacement for Weylets but rather as an
alternative in the framework of the projected von Neumann basis.}

\subsection{Numerical implementation}
\label{ch:pW_numImpl}
Since the set of used basis functions changes during pruned {time-dependent} dynamics, a careful implementation is required for obtaining an efficient algorithm. Notably, the tensor transformation needs special attention because this is the main bottleneck of the dynamics. A very efficient implementation optimized for operators with \SOP form is given in the appendix. 
It takes advantage of the fact that the algorithm simplifies significantly if the tensor transformation is performed over the last dimension (assuming this dimension is represented contiguously in memory). For other dimensions, the coefficient tensor is first permuted properly which is not costly compared to the tensor transformation itself.

If a DVR basis is used, no \SOP form of the potential is needed and the multiplication of the potential times the vector is just a (direct) vector-vector multiplication, see section \ref{sec:theory_overview}. We exploit this for exact dynamics without pruning. For our tests with pruned DVRs, we use the \SOP form, for convenience. The potential is then represented as a sum of product of diagonal matrices. This introduces an overhead compared to a simple vector-vector multiplication. 

The decision to add or remove basis functions is defined by a wave amplitude
threshold, $\theta$. If the absolute value of a coefficient is larger than or equal
to $\theta$, the nearest neighbors of the corresponding basis in phase space
are added to the set of used basis functions. If they have not already been
members of the set of used basis functions, they are included with coefficients set to zero.
If the value is smaller than $\theta$ and all nearest neighbors have
coefficients whose absolute value is smaller as well, the corresponding basis
function is excluded from the set. Hartke used two thresholds, one for the
inclusion and one for the exclusion of basis
functions.\cite{proDG_hartke_2006,proDG_hartke_2008} We decided to choose only
one threshold but we add the possibility to exclude basis functions only if the total number of those to exclude is larger than a certain relative threshold.\cite{pvb_edyn_tannor_2015,pvb_algorithms_tannor_2016}
It remains to choose \emph{how many} nearest neighbors are added. This is tested numerically in section \ref{ch:dwell}.

The algorithm and the choice of data structures needed to check the
coefficients are crucial. In our implementation, we simply loop over all basis
function coefficients and evaluate them successively. The new set of basis
function indices are stored in a hash table.\footnote{We use
  \texttt{std::unordered\_set} of the C++ programming language standard
  library as implemented in the GNU compiler collection\cite{gcc_manual}}\textsuperscript{,}\cite{sedgewick_book}
If new basis functions
need to be added, a lookup is required to check whether these basis functions
are already elements of the set. If not, they are added to the set. The usage
of the hash table is very important because lookup and insertion scales on
average as $\mathcal O(1)$. Since all basis functions have to be checked, the
whole adaption procedure scales as $\mathcal O\left(\nMean{\widetilde{n}}^{D}\times
  N_\text{neighbor}\right)$, where $N_\text{neighbor}$ is the number of
nearest neighbors of one basis function. This number is much smaller than
$\nMean{\widetilde{n}}^{D}$, see section \ref{ch:dwell} for a test of how many
neighbors have to be added. If no hash table but a simple sorted list of
coefficients were used, the scaling would be $\mathcal O\left(\nMean{\tilde
    n}^{D}\times  N_\text{neighbor}\times\nMean{\widetilde{n}}^{D}\right) =
\mathcal O\left(\nMean{\widetilde{n}}^{2D}\times N_\text{neighbor}\right)$, because
insertion and removal of duplicates in a sorted array scales linearly. The
adaption would then scale worse than the matrix-vector product and would hence
become the computational bottleneck of the dynamics. With our implementation
using hash tables, the adaption of the set of basis functions never needs 
more than 5 to 10\% of the overall computing time during the dynamics,
in the application examples shown below. It could be further optimized by storing whether a basis function has neighbors with large coefficient values.\cite{proDG_hartke_2008} If that is the case, there is no need to check or to add neighbors to this basis function.

\section{Application}
\label{sec:appl}
\subsection{Two-dimensional double well}
\label{ch:dwell}
We test the {time-dependent} dynamics within a two-dimensional double well employed in \lit{pvb_math_tannor_2016}. The Hamiltonian for this model potential is
\begin{equation}
\begin{split}
 \hat H_\text{DW} =& -\frac1{2\times 200}\left(\partdd{}{x}+\partdd{}y\right)+ 6.4 (x-1)^2\times\\
 &(x-2)^2 + 37.5 (y-2)^2 + 10  x^2 y,\label{eq:ex_dwell}
\end{split}
\end{equation}
and the used initial wave packet has the form
\begin{equation}
\begin{split}
 \braket{xy}{\Psi(0)} =& \frac{\sqrt{2/\pi}}{(0.0008)^{1/4}}\exp[-(x-2.1)^2/0.04 -\\
 &(y-2.05)^2 / 0.02].
 \label{eq:ex_dwell_psiInitial}
 \end{split}
\end{equation}
The wave packet is propagated until $t_\text{e}=24.6$. In $x$ we use a phase-space grid of size $n_x\times n_p = 15\times 11$ ($x\in [-0.5,3.7]$), and in $y$ a grid of size $9\times 9$ ($y\in[1.0,2.9]$). The overall size of the basis is $165\times 81 = 13365$. We use the FGH method  for the pruned dynamics and as the underlying DVR of the projected Weylets and \pvb. For further details, we
refer to \lit{pvb_math_tannor_2016}. We employ the short iterative Arnoldi
propagator\cite{sil_light_1986,mctdh_rev_meyer_2000} as implemented in the
Heidelberg MCTDH package.\cite{mctdh_package}

To evaluate the dynamics, we compute the autocorrelation, $C(t)$, using
\begin{equation}
 C(t) = \braket{\Psi(0)}{\Psi(t)} = \braket{\conj{\Psi(t/2)}}{\Psi(t/2)},\label{eq:acorr}
\end{equation}
where the last equality holds for real-valued initial wave functions $\ket{\Psi(0)}$.\cite{acorr_2t_engel_1992,NO2_mctdh_cederbaum_1992} To compute the autocorrelation with a nonorthogonal basis like \pvb, one needs to multiply the coefficient vector $\vec a$ with $\matr B^T\matr B$, where $\matr B$ is defined in \eq\eqref{eq:pvb_H_trafo}. For $D=1$, this is
\begin{equation}
\!\! C(2t) = \braket{\conj{\Psi(t)}\!}{\Psi(t)} = \conj{\left(\vec a^\dagger \matr B^\dagger\right)} \matr B \vec a = \vec a^T \matr B^T \matr B \vec a.
\end{equation}

From section \ref{ch:pW_numImpl}, it remains to show how many nearest
neighbors, $N_\text{neighbor}$, should be added during the adaption of the
pruned basis. At first attempt, one could add all nearest neighbors which are
directly connected to the basis function of interest, \ie also basis functions
located diagonally on the phase-space grid. This means an index change of the
basis of $\pm 1$ in \emph{each} dimension in phase space and all possible
combinations of index changes. This corresponds to a ``phase-space ball radius'' of $2$.\cite{pvb_algorithms_tannor_2016} Obviously, this is the most conservative way to add basis functions but it requires to add many basis functions in high dimension since the number of all nearest neighbors scales again exponentially, namely $3^{2D}-1$. The factor of $2$ in the exponent comes from the definition in phase space and not in coordinate space. For high-dimensional problems like 24-dimensional pyrazine,\cite{pyrazine_24d_cederbaum_1998} this number is $3^{48}$ which approximately equals Avogadro's number. A much simpler way would be to exclude all diagonally connected neighbors and add a maximum of two basis functions \emph{per dimension}, \ie only $2D$. This corresponds to a phase-space ball radius of $1$ and was chosen in \lit{proDG_hartke_2008}. Clearly, this way works even for high-dimensional problems but the basis adapts less quickly during the dynamics, which maybe critical for tunneling processes etc. On the other hand, this could be compensated with a smaller threshold but then, overall more basis functions may be required.
An intermediate way would correspond to a phase-space radius of $\sqrt{2}$, where $2D^2$ nearest neighbors are added.\cite{pvb_algorithms_tannor_2016} 

\begin{figure}
\includegraphics{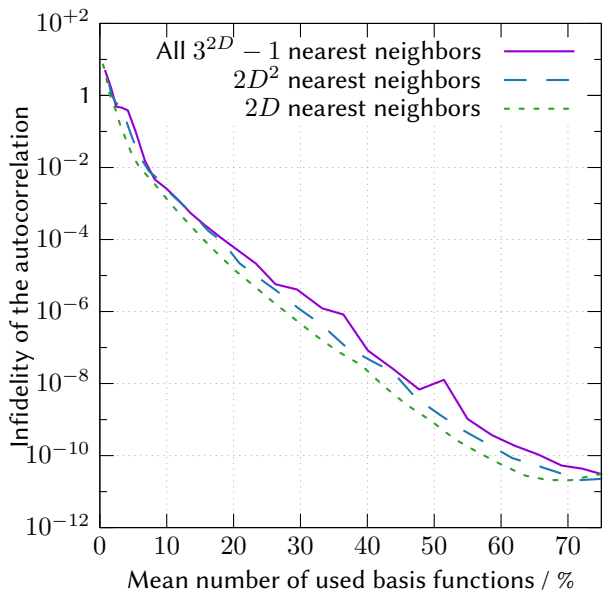}
\caption{Accuracy of the pruned projected Weylets dynamics for the two-dimensional double well as a function of the percentage ratio of reduced and unreduced basis sizes. {The full basis size is 13365}.
The accuracy is determined by the infidelity of the autocorrelation and shown for three different numbers of nearest neighbors to add to the pruned basis (see the text for details).}
 \label{fig:dwell_ballRadComp}
\end{figure}

\Fig \ref{fig:dwell_ballRadComp} compares the three choices for the double-well dynamics. We have evaluated the infidelity in the autocorrelation,
defined by the Euclidian $L_2$ distance between the values of $C(t)$ of the
pruned and of the unpruned, exact dynamics. We have used different wave
amplitude thresholds but plot the mean number of used basis functions (in
percent compared to the unpruned dynamics). The smaller the threshold, the
more basis functions are added and the larger the number of used basis
functions. Because the map of threshold to number of basis functions is different
for each problem and depends on the dimensionality and the number of basis
functions to add, it makes more sense to plot the number of actually used
basis functions than the threshold employed.
{For the double well and \ce{NO2} computations (section \ref{sec:NO2}), the threshold has been varied between $0.046$ and $10^{-12}$.}
Our results show that the best
way is {always to} add only $2D$ nearest neighbors. Then, fewer basis functions
are {required} for the same accuracy, although a smaller threshold is needed. 
We did the same test for the \ce{NO2} dynamics (section \ref{sec:NO2}) and came to the same conclusion.
{Henceforth, we always add only $2D$ nearest neighbors (phase-space ball radius of $1$) to each pruned basis function.}

\begin{figure}
\includegraphics{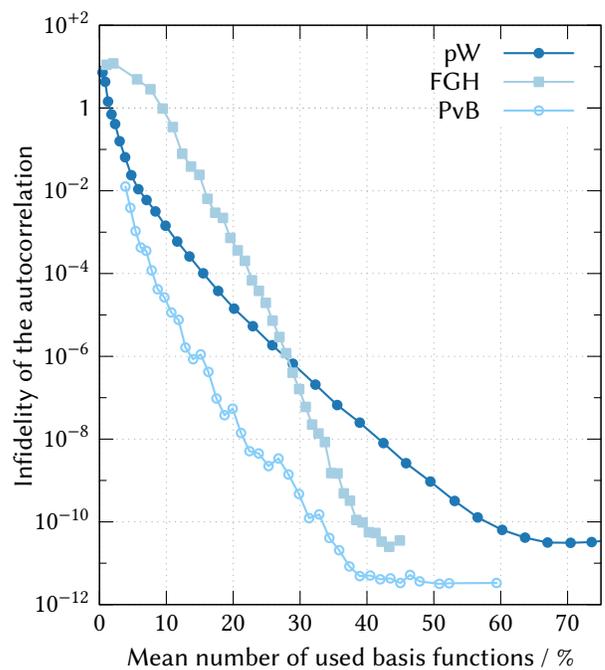}
\caption{Accuracy of the dynamics for the 2D double well as a function of the percentage ratio of reduced and unreduced basis sizes. {The full basis size is 13365}.
The accuracy is determined by the infidelity of the autocorrelation and shown for the projected Weylets (\pW, filled circles), pruned FGH (squares) and \pvb (rings).}
 \label{fig:dwell_eval_nBas}
\end{figure}

The accuracy of the pruned dynamics as a function of the number of used basis functions is shown in \fig\ref{fig:dwell_eval_nBas} for the different methods tested.
The accuracy is evaluated by comparing to an exact unpruned FGH basis. For the pruned bases, we have compared a FGH basis that is pruned in coordinate space, the projected Weylets (\pW) and the biorthogonal projected von Neumann basis (\pvb).
The accuracy is measured by the error in the autocorrelation. We have also compared the error of the final wave packet. The curves for this error measurement look very similar such that we only show the error in the autocorrelation. 
All computations have been performed using a single core of Intel(R) Xeon(R) CPU E5-2650 v2 processors. 
The accuracy of the integrator is adapted to the wave amplitude threshold.
\pvb{} {requires the fewest basis functions for a given accuracy.} {\pW in general requires more basis functions than \pvb to represent a state. However, momentum-asymmetric states are described by \pW just as well as momentum-symmetrized states (see section {iii} in the supplementary information for an example).}

{The different accuracy limits stem from the numerical error in the creation of the bases (arising from multiplication by $\matr S^{-1/2}$ or $\matr S^{-1}$). 
In principle, both \FGH, \pW and \pvb should give the same result if $100\%$ of basis functions are used. 
The residual error on the order of $10^{-10}$ in the fidelity of the autocorrelation arises from the accumulation of numerical noise which we have not tried to reduce.}

For low accuracies, the pruned FGH basis needs many more basis functions than \pvb or \pW to obtain the same accuracy. Surprisingly, FGH \emph{beats} \pW for errors smaller than $10^{-6}$. In fact, the number of basis functions required to reach numerically exact results is similar to that in \pvb, namely $\sim\unit[40]\%$ of the totally available basis. This comes from the choice of potential and wavepacket. The potential is quite correlated between $x$ and $y$, and in certain regions the whole phase space needs to be covered. Then, it is sufficient to use a DVR basis that is not localized in phase space. 
Note that a FGH basis is more localized in $x$ than a phase-space-localized basis (here, it is a factor of $\sim10$ times more localized in $x$ than the \pW basis). However, we could change the ratio $\Delta x/\Delta p$ for the phase-space-localized functions to make them more localized in $x$ and even use more complicated phase-space tiling.\cite{pvb_wavelet_tannor_2012} We prefer not to optimize the tiling for this problem to keep the discussion more general. 

Another point to consider is that some newly added nearest neighbors do not contribute to the wavepacket if it moves into another direction. This is more severe in phase space where there are four directions to consider instead of two in coordinate space. In fact, the current scheme is wasteful for phase-space dynamics. For small wave amplitude thresholds $\theta$ (high accuracy), about $\unit[20]\%$ of basis functions whose coefficient value is smaller than $\theta$ are used, \ie they do not contribute to the description of the wavepacket but are included because a nearest neighbor has large coefficient values. For larger $\theta$, this value increases even up to $\sim\unit[50]\%$. As expected, the situation is less severe for the DVR that is only localized in coordinate space. There, only $\sim\unit[10]\%$ of functions are ``wasted''. This could be corrected by better schemes to add nearest neighbors. Essentially, the information where the wavepacket moves is already encoded in the phase-space structure. 

\begin{figure}
\includegraphics{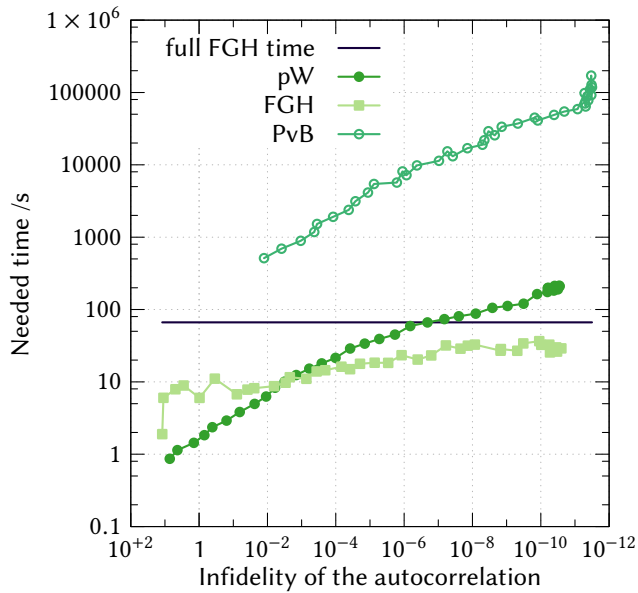}
\caption{Computing time of the dynamics for the 2D double well as a function of accuracy. The accuracy is determined by the infidelity of the autocorrelation and shown for the projected Weylets (\pW, filled circles), pruned FGH (squares) and \pvb (rings). The black horizontal line denotes the computing time of the unpruned FGH method.}
 \label{fig:dwell_eval_time}
\end{figure}

We now compare the timing. \Fig\ref{fig:dwell_eval_time} compares the accuracy versus {computing} time for the pruned dynamics. Due to the $\mathcal O(\nMean{\widetilde{n}}^{2D})$
scaling, \pvb requires more than $600$ times more {computing} time than any other
method. Even though FGH needs more basis functions to reach the same accuracy,
the pruned dynamics is almost always as fast or faster than \pW. {Only at the low-infidelity end is \pW faster.}
Both FGH and \pW use
a \SOP Hamiltonian and employ the same computational routines. The difference
comes from the transformation of the Hamiltonian, see \eq
\eqref{eq:pvb_H_trafo}. For a DVR basis like the FGH method, all potential
matrices are diagonal whereas in \pW, the matrices are nondiagonal although
sparse (actually banded if seen as a tensor in phase space). A multiplication of a diagonal matrix times the coefficient vector scales as $\mathcal O\left(\nMean{\widetilde{n}}^{D}\right)$ instead of $\mathcal O\left(\nMean{\widetilde{n}}^{D+1}\right)$.

\subsection{\ce{NO2} dynamics on the $B_2$ surface}
\label{sec:NO2}
As a three-dimensional example, we study the performance of our methods with
the dynamics of \ce{NO2} on the $B_2$ surface. Due to its ergodicity, this
dynamic is challenging for the MCTDH
method.\cite{NO2_mctdh_cederbaum_1992,correlated_vN_entropy_effective_Hamiltonian_manthe_2012}
The wave packet spreads over many configurations in phase space and large
basis sets are required. Nevertheless, as we show here, even in such a
situation pruning is useful, since the wavepacket does not instantly and fully
cover the full space spanned by a direct-product basis.

We follow \lit{NO2_mctdh_cederbaum_1992} and propagate the wave function in bond coordinates (distances $r_1$ and $r_2$ for \ce{N-O} and bond angle $\theta$). In these coordinates, the vibrational Hamiltonian takes the form of
\begin{widetext}
\begin{align}
 \hat H_{\ce{NO2}} =& -\frac1{2\mu} \left(\partdd{}{r_1} + \partdd{}{r_2}\right) - \frac{\cos(\theta)}{m_{\ce N}} \frac{\partial^2}{\partial r_2 \partial r_1} -\frac{1}{2\mu} \left(\frac1{r_1^2}+\frac1{r_2^2}\right) \frac1{\sin(\theta)} \partd{}{\theta}\nonumber\\%
 &+%
 \frac1{2m_{\ce{N}}r_1r_2}\left[\cos(\theta),\frac1{\sin(\theta)}\partd{}{\theta}\sin(\theta)\partd{}{\theta}\right]_{+}+
 \frac{1}{m_{\ce{N}}}\left(\frac1{r_1}\partd{}{r_2}+\frac1{r_2}\partd{}{r_1}\right)\partd{}{\theta}\sin(\theta)\\
 &+ V(r_1,r_2,\theta),\nonumber
\end{align}
\end{widetext}
with
\begin{equation}
 \mu^{-1} = m_{\ce{O}}^{-1} + m_{\ce{N}}^{-1},
\end{equation}
where $m_{\ce{A}}$ is the mass of atom $\ce{A}$. $[\cdot,\cdot]_+$ denotes the anticommutator. The potentials $V(r_1,r_2,\theta)$ for the $A_1$ and the $B_2$ states are taken from \lit{no2_pes_petrongolo_1991} and the latter modified as explained in \lit{NO2_mctdh_cederbaum_1992}.
To compensate for inaccuracies of the potential,\cite{no2_pes_buenker_1994} the potential functions are held constant below $r_i<1.95$ and below $\theta < 1$.
The volume element is
\begin{equation}
\dd V = \dd r_1 \dd r_2 \dd \theta \sin(\theta).
\end{equation}

The projected von Neumann basis has already been used in angular coordinates (Jacobi and Radau) with a Gauss-Legendre DVR in \lit{pvb_LiCN_tannor_2014,pvb_H2O_carrington_2015}. The used grid points are $\theta_i = \arccos(z_i),\ \theta_i\in[0,\pi]$, where $z_i\in [-1,1]$ is a Gauss-Legendre DVR point. The grid points $\theta_i$ are almost equidistantly spaced and can thus be used just like a FGH grid. The condition numbers of $\matr G$ (\pvb) or $\matrgreek \Theta$ (projected Weylets) are less than $20$ for a Gauss-Legendre grid with $100$ grid points. A projected phase-space basis is therefore well conditioned for angular grids.\footnote{Brown and Carrington noted problems with the condition number of $\matr G$, namely numbers $\sim 10^3$ after an adaption of the width of each von Neumann function for a grid of size $48$.\cite{pvb_H2O_carrington_2015} We do not experience such problems.}

We use a FGH DVR basis of size $n_x\times n_p = 13\times 13$ in the radial coordinates ($r_i\in[1.6,16]$) and a Gauss-Legendre DVR of size $10\times 10$. In total, the unpruned basis has a size of $\sim 285\times 10^4$. The ground state of the $A_1$ surface is taken as the initial wave packet and propagated on the $B_2$ surface until $t_\text{e}=\unit[90]{fs}$.  The autocorrelation, \eq\eqref{eq:acorr}, is computed to evaluate the pruned dynamics.

\begin{figure}
\includegraphics{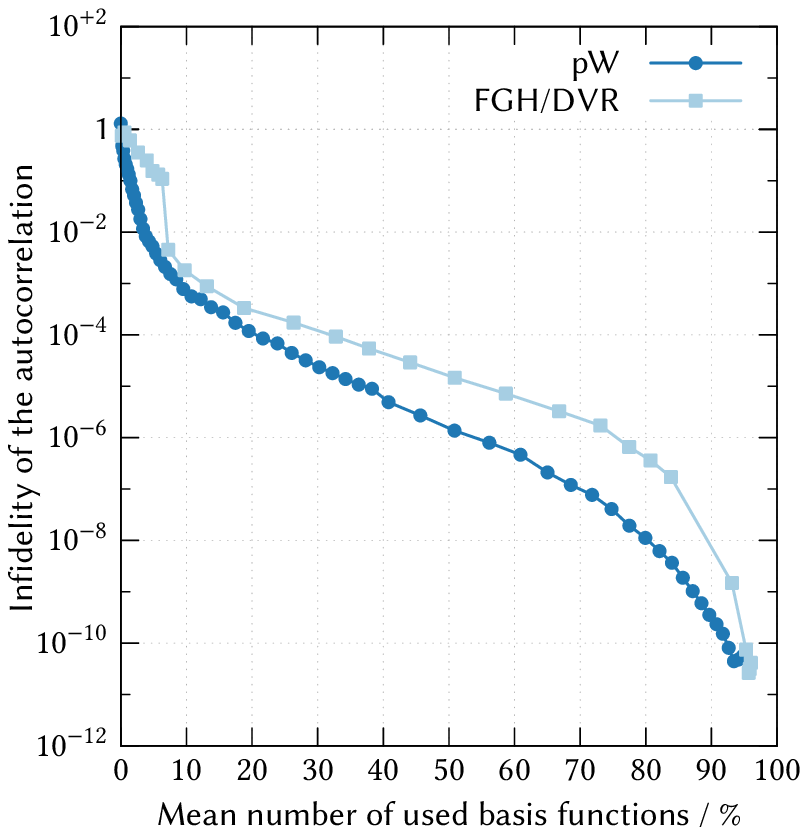}
\caption{Accuracy of the dynamics for \ce{NO2} as a function of the percentage ratio of reduced and unreduced basis sizes for the pruned Weylets (circles) and pruned FGH/DVR (squares), compare with \fig\ref{fig:dwell_eval_nBas}. {The full basis size is $\sim285\times 10^4$.}
}
 \label{fig:NO2_eval_nBas}
\end{figure}

The accuracy of the pruned dynamics is depicted in \fig
\ref{fig:NO2_eval_nBas}. We are unable to show results for \pvb due to the
demanding requirements in computing time. From section \ref{ch:dwell}, it
should be clear that pruned orthogonal bases are faster than both \pvb and pruned nonorthogonal bases, in general. \pW always needs fewer basis functions than a pruned FGH/DVR basis to
reach the same accuracy. Almost all basis functions
are required to reach numerical accuracy because the dynamics is highly
ergodic and most of phase space is covered. 

\begin{figure}
\includegraphics{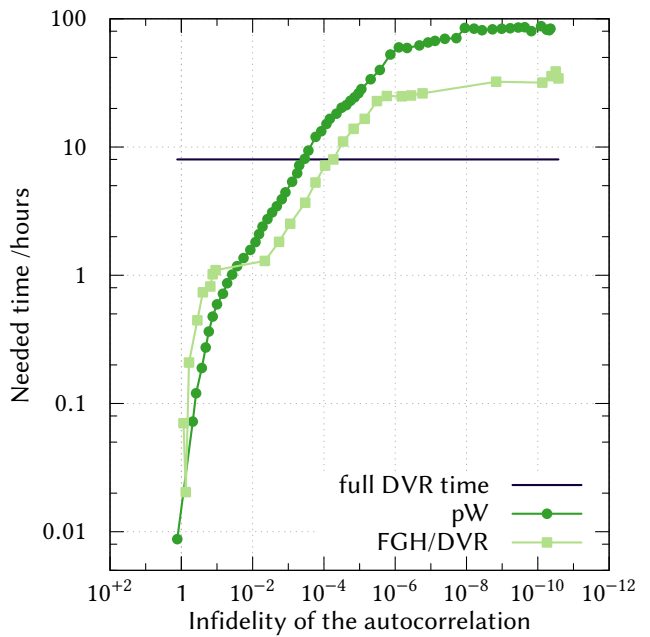}
\caption{%
Computing time of the dynamics for \ce{NO2} as a function of the accuracy for the pruned Weylets (circles) and pruned {FGH/}DVR (squares), compare with \fig\ref{fig:dwell_eval_time}.
The black horizontal line denotes the computing time of the unpruned {FGH/}DVR method. Due to computational overhead, pruned DVR dynamics needs more computing time than full DVR dynamics, for many basis functions (small infidelities). To some extend, this is artificial due to the choice of implementation, see section \ref{ch:pW_numImpl} for details.}
 \label{fig:NO2_eval_time}
\end{figure}

\Fig \ref{fig:NO2_eval_time} shows the timing of the \ce{NO2} dynamics.
Like in the double-well example, pruned DVR {almost} always performs faster than \pW. For accuracies lower than $7\times10^{-5}$ (less than $\unit[38]\%$ of basis functions), pruned DVR dynamics is faster than the exact unpruned dynamics. The pruned DVR simulations could be accelerated by a better exploitation of the diagonality of the potential. This is done for the full DVR but not for the pruned DVR, see section \ref{ch:pW_numImpl}.
To reach an accuracy of $10^{-4}$, DVR needs $5$ hours computing time whereas \pW requires $12$ hours. 
However, DVR also needs more basis functions to reach the same accuracy. For the considered accuracy, $\unit[26]\%$ of basis functions are required for the DVR dynamics whereas only $\unit[16]\%$ are required for \pW. 
The memory requirements are not of interest in this three-dimensional example because no more than $\sim\unit[1]{GB}$ of memory is ever needed. However, for higher-dimensional dynamics as ergodic as those of \ce{NO2}, memory could be crucial and \pW might be the method of choice.

\begin{figure}
\includegraphics{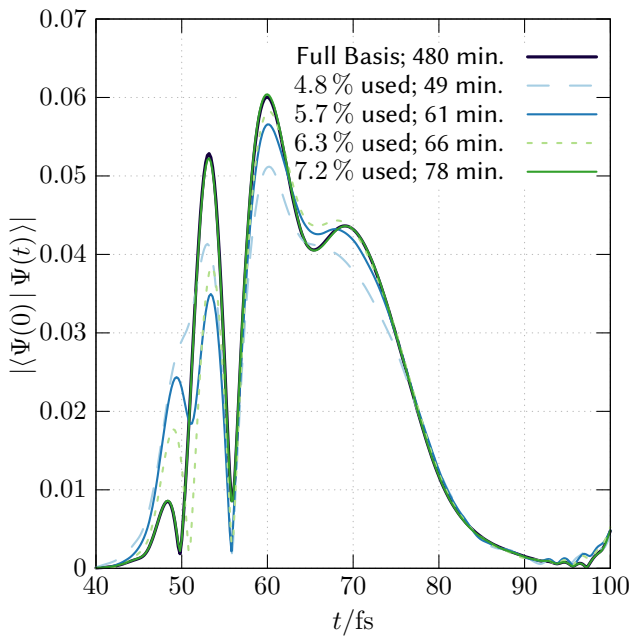}
\caption{Absolute value of the autocorrelation for pruned {FGH/}DVR dynamics compared to the exact dynamics (black line) for \ce{NO2}.}
 \label{fig:NO2_acorr}
\end{figure}

Comparing the error in the autocorrelation to very high accuracies is a useful 
benchmark of pruned methods. However, for many purposes even a lower accuracy autocorrelation function is sufficient to obtain the correct qualitative behavior of the spectral observables.
The autocorrelations for some simulations are shown in
\fig \ref{fig:NO2_acorr}. There, we only compare the DVR dynamics
because this outperforms \pW. Most of the features are reproduced even if only
$\unit[4.8]\%$ of the total number of available basis functions are used during the
dynamics. The autocorrelation is visually converged if $\unit[7.2]\%$ are used
and the dynamics is faster by a factor of $6$ compared to the full dynamics.

\subsection{Nonadiabatic dynamics of pyrazine}
As a higher-dimensional example, we consider the nonadiabatic dynamics of a six-dimensional vibronic-coupling model of pyrazine.\cite{
pyrazine_24d_cederbaum_1998} The Hamiltonian was obtained from the Heidelberg MCTDH package.\cite{mctdh_package}
Compared to the dynamics of \ce{NO2}, the wavepacket behaves much more smoothly. Hence, many fewer basis functions are needed and the dynamics is well suited for the MCTDH method. For further details of this benchmark example, we refer to the literature.\cite{pyrazine_24d_cederbaum_1998,pyrazine_24d_cederbaum_1999}

Again, we used a FGH basis with parameters shown in
\tab\ref{tab:pyr6_basis}. In this case, a Gauss-Hermite DVR would be better
suited but to allow for an easier comparison to the other examples and to the
projected Weylet method, we choose to keep the FGH method.
The overall basis size, including the electronic basis consisting of two states, is $\sim 322\times 10^6$. We have propagated until $\unit[95]{fs}$.
\begin{table}
\caption{Used basis parameters of the six-dimensional pyrazine example.}
\label{tab:pyr6_basis}
\begin{ruledtabular}
 \begin{tabular}{lccc}
  Mode & $n_x$ & $n_p$ & $x$-Range\\\hline
  $\nu_{10a}$ & $5$ & $6$ & $[-8.3,8.3]$ \\
  $\nu_{6a}$ & $6$ & $7$ & $[-9.4,9.4]$\\
  $\nu_1$ & $5$ & $5$ & $[-7.0,7.0]$\\
  $\nu_{9a}$ & $5$ & $4$ & $[-7.1,7.1]$\\
  $\nu_{16b}$ & $4$ & $4$ &$[-6.6,6.6]$\\
  $\nu_{18b}$ &  $4$ & $4$ &$[-6.6,6.6]$\\
 \end{tabular}
 \end{ruledtabular}
\end{table}

\begin{figure}
\includegraphics{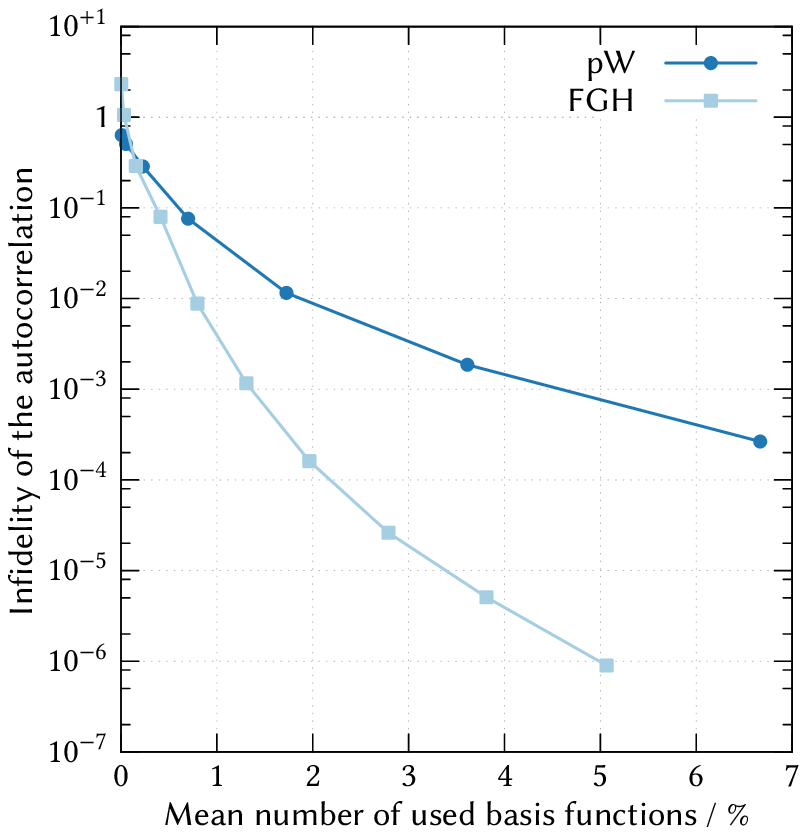}
\caption{Accuracy of the dynamics for the six-dimensional model of pyrazine  as a function of the percentage ratio of reduced and unreduced basis sizes for the pruned Weylets (circles) and pruned FGH (squares), compare with \fig\ref{fig:dwell_eval_nBas}. {The full basis size is $\sim322\times 10^6$.}}
 \label{fig:pyr6_eval_nBas}
\end{figure}

The accuracy versus number of used basis functions is depicted in \fig \ref{fig:pyr6_eval_nBas}.
{The wave amplitude threshold $\theta$ has been varied between $10^{-3}$ and $10^{-8}$.}
Only \unit[5]\% of the totally available FGH basis is needed to accurately describe the dynamics. This comes primarily from the smooth dynamics where the wavepacket retains a rather compact form in configuration space. However, it also shows the general trend that more and more basis functions are wasted if direct product bases are used in higher-dimensional spaces.\cite{nonprod_grid_carrington_2009,proDG_hartke_2008}

Surprisingly, the FGH method needs \emph{fewer} basis functions than the
projected Weylets for reaching the same accuracy. Obviously, localization in
coordinate space is here much more important than phase-space localization. In
general, fewer basis functions are needed in each mode and the wavepacket does
not reach high momenta. Further, the needed ``shell'' of nearest neighbors is
more wasteful in phase space than in coordinate space. Consider the two modes
with a size of $n_x\times n_p=4\times 4$. In coordinate space, only two
nearest neighbors are added in one dimension to one function, that is $1/8$ of
basis functions are added. In phase space, four nearest neighbors need to be added which is already $1/4$. In other words, a $4\times4$ grid in phase space is simply too small with the current scheme.
Nevertheless, adding nearest neighbors also here is wasteful for the DVR method. During the dynamics, about $\unit[50]\%$ of included basis functions have values smaller than the used threshold (compare with the discussion in section \ref{ch:dwell}). For \pW, it is $\unit[80]\%$.
Note that, for slightly different reasons, Halverson and Poirier have found that a properly pruned harmonic oscillator basis can be more efficient than the symmetrized Gaussians in certain regions of the vibrational spectrum.\cite{benzene_hybrid_truncation_scheme_HO_basis_poirier_2015}
Brown and Carrington have found similar results.\cite{symmetrized_gaussians_sop_carrington_2016}

\begin{figure}
\includegraphics{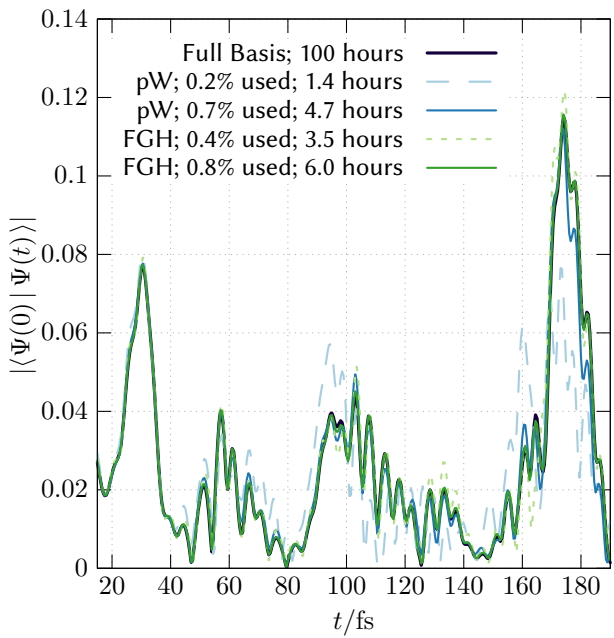}
\caption{Absolute value of the autocorrelation for pruned projected Weylets and FGH dynamics compared to the exact dynamics (black line) for the six dimensional model of pyrazine.}
 \label{fig:pyr6_acorr}
\end{figure}

The convergence of the autocorrelation function for selected simulations is shown in \fig \ref{fig:pyr6_acorr}. Only $\unit[0.8]\%$ of basis functions are needed for visual convergence of the autocorrelation! The computational speed-up compared to exact dynamics is a factor of more than $16$. Note that the set-up of the permutations and arrays $\mathcal I$ and $\mathcal J(i)$ (see the appendix for details) needed for an efficient matrix-vector product took significantly more computing time than before, namely $\sim\unit[30]\%$  for the largest computation. This fraction is negligible for our two- and three-dimensional examples where, in general, many fewer basis functions are needed. The reason for the increased computing time is that this part is currently not optimized in our implementation. Hence, the computing time could be significantly reduced.

\section{Conclusion and outlook}
\label{sec:conclusion}
By combining the idea behind the biorthogonal von Neumann basis (\pvb), namely projecting phase-space-localized von Neumann functions to another basis, with the orthogonalized, momentum-symmetrized Gaussians (Weylets), we have established a new method, projected Weylets (\pW), that inherits the advantage of \pvb, namely DVR accuracy, and the advantage of Weylets, orthogonality. We have shown that orthogonality is crucial for an efficient matrix-vector product for solving the time-dependent Schrödinger equation of multidimensional systems. Additionally, we have developed a new, highly efficient algorithm for the matrix-vector product for sum-of-product Hamiltonians by permuting the basis such that the dimension to transform over is contiguously represented in memory. 

We have compared three methods for dynamically pruning a localized basis during the wavepacket dynamics, namely \pvb and \pW that are localized in phase space and the FGH and Gauss-Legendre DVRs that are localized in coordinate space. Due to the nonorthogonality and the unfavorable scaling of the matrix-vector product, \pvb is much more expensive than the other methods, although it is very appealing in theory.
It may, however, be a very useful tool for one-dimensional simulations, since the representation in \pvb is the most compact one. By examining the ergodic dynamics of \ce{NO2}, we showed that, due to the phase-space localization of \pW, it needs fewer basis functions than a pruned FGH basis for the same accuracy. However, we showed further that in general \pW needs more computing time because all potential matrices are non-diagonal whereas they are diagonal in the DVR representation. For the six-dimensional dynamics of a vibronic coupling model of pyrazine, coordinate-space localization turned out to be much more important than phase-space localization. There, FGH outperforms \pW not only in computing time but also in accuracy versus number of basis functions. Fewer than $\unit[1]\%$ of all basis functions were needed to obtain almost converged dynamics, resulting in speedups of more than $16$.
Only a single parameter is needed to control the accuracy of the simulation.

At first sight, DVR methods seem to have more advantages than \pW. The
potential terms are diagonal, and in general no sum-of-product form of the
potential operator is required. The \pW method gives, however, more compact
representations of the wavefunction if the latter covers many regions in phase
space and many basis functions are needed, like for the \ce{NO2} dynamics. A
compact representation is very crucial for higher-dimensional dynamics. 
This might be especially useful for scattering simulations like \ce{S+ + H2} or \ce{HO + CO}.\cite{S+_H2_scattering_bulut_2016,HO_CO_dyn_guo_2012}
Note further that \pW and a pruned DVR can of course be combined: \pW can be used in modes where phase-space localization is needed and the pruned DVR can be used in spectator modes that are more well-behaved.

Our methods can, of course, be improved. At this stage, we add nearest neighbors in all directions, even if the wave packet moves only in one direction. This is wasteful, especially in phase space. The phase-space representation of \pW can be utilized to predict the movement of the wavepacket. By this, it is clear where to add nearest neighbors. In this contribution, we made use of a sum-of-product Hamiltonians. A transformed Hamiltonian of general form should, however, be very sparse in the \pW representation. This could lead to a very promising tool to handle high-energy electronic dynamics, like double ionization dynamics. There, the potential of interest (Coulomb) is not decomposable. Exploiting the sparsity could be useful even if many terms are needed for the sum-of-product potentials. 
The phase-space view of \pW and \pvb offer further possibilities to reduce the size of the basis, either by contraction of the basis in the classically forbidden region or by combination with semiclassical methods.\cite{asaf_thesis} 
Note that pruned dynamics also offers the possibility to compute the potential on the fly, obviating complicated global representations of the potential energy surface. These avenues will be pursued in the future. 

Why bother with pruned dynamics if (ML-)MCTDH is the method of choice for
high-dimensional dynamics? First of all, MCTDH has its drawbacks: it is very expensive for highly correlated systems and is inefficient if many primitive basis functions are required. But perhaps more significantly, pruned dynamics, either in coordinate space or phase space, can be combined with MCTDH, in different possible ways. One way is to use pruned functions as the underlying basis of the single-particle functions (SPFs) in MCTDH, which is quite similar to the multilayer variant of G-MCTDH.\cite{G-MCTDH_layer_Burghardt_2013} A second way is to prune the MCTDH coefficient tensor by transforming the SPFs to a localized basis.\cite{mctdh_selected_configurations_worth_2000} Both ways can be easily combined. Work in this direction is in progress.

\section*{Supplementary material}
{
  See supplementary material for {more details on the transformation of the Weylet basis compared to \pW,} further comparisons of Weylets against \FGH, \pvb and \pW and for a comparison of a phase-space-representation of a momentum-symmetric and a momentum-asymmetric state.}

\begin{acknowledgments}
H.~R.~L.~and D.~J.~T.~thank S.~Machnes and E.~Assémat for enlightening discussions regarding the \pvb method.
We thank H.~D.~Meyer for providing us with the Heidelberg MCTDH package and U.~Manthe for help with the \ce{NO2} potential.
We thank T.~Carrington, {B.~Poirier and the anonymous Referees} for helpful comments on the manuscript.
H.~R.~L.~acknowledges support by the Fonds der Chemischen Industrie,
Studienstiftung des deutschen Volkes, and the Deutscher Akademischer Austauschdienst.
D.~J.~T.~acknowledges support from the Israel Science Foundation (533/12) and from the Minerva Foundation with funding from the Federal German Ministry for Education and Research.
\end{acknowledgments}

\appendix*
 
\section{Tensor transformations}
\label{app:matric_vector_product}
Let us first consider the tensor transformation (matrix-vector product) without pruning. By doing the transformation sequentially, one always has to consider transformations in only one dimension $t$,
\begin{equation}
 \tilde a_{i_1i_2\dots i_D} = \sum_{j_t=1}^{n_t} h^{(t)}_{i_tj_t} a_{i_1\dots i_{t-1}j_ti_{t+1}\dots i_D}.\label{eq:tens_trafo_3d}
\end{equation}
To simplify the notation, we define the multiindices $i\equiv i_1\dots i_{t-1}$ and $k\equiv i_{t+1}\dots i_D$. We further denote the $t$th index as $j$ or $l$ and drop the superscript $(t)$ of the matrix. Then, \eq\eqref{eq:tens_trafo_3d} simplifies to 
\begin{equation}
 \tilde a_{ilk} = \sum_{j=1}^{n_t} h_{lj} a_{ijk}.
\end{equation}
Assuming a row-major layout of the tensor (the last index is contiguous in memory), a possible implementation would look like
\begin{lstlisting}
$a'_{:::} \leftarrow 0$
for $i$ in $[1,n_1]$:
  for $l$ in $[1,n_2]$:
    for $j$ in $[1,n_2]$:
      for $k$ in $[1,n_3]$:
        $a'_{ilk} \leftarrow a'_{ilk} + h_{lj} a_{ijk}$
\end{lstlisting}
A colon means an implicit loop over all indices.
The order of the loops is crucial to enable an efficient caching of the values
in memory. Actually, this transformation can be recast into a loop over $n_1$ general matrix matrix multiplications (GEMM) \emph{without} additional copying:\cite{tensor_contractions_BLAS_usage_bientinesi_2014}
\begin{lstlisting}
for $i$ in $[1,n_1]$:
  $a'_{i::} \leftarrow \matr{h} \matr{a}_{i::}$
\end{lstlisting}
where $\matr{a}_{i::}$ is a $n_2\times n_3$ matrix. Because there are many efficient libraries implementing the matrix matrix product, 
this recasting is favorable, although, in real applications, a speed up of only up to $\sim 10\%$ is achieved because $\matr{a}_{i::}$ are long and skinny since $n_2 \ll n_3$. Recall that $n_3$ is the number of basis functions of a multiindex and therefore very large. Memory access is thus the bottleneck of the tensor transformation -- even if well tuned algorithms are employed for the matrix matrix product.\cite{gemm_goto_2008}

For a pruned basis, the situation is much more complicated. Following the definitions introduced in section \ref{ch:pvb_drawbacks}, a straightforward implementation would look like
\begin{lstlisting}
$a'_{:::} \leftarrow 0$
for $i$ in $\mathcal I$:
  for $l$ in $\mathcal J(i)$:
    for $j$ in $\mathcal J(i)$:
      for $k$ in $\mathcal K'(i,j,l)$:
        $a'_{ilk} \leftarrow a'_{ilk} + h_{lj} a_{ijk}$
\end{lstlisting}
The last loop over $k$ depends on $\mathcal K'(i,j,l)$ since one has to sum \emph{only} over entries where $k$ is element of both sets $\mathcal K(i,l)$ (left side of the summation) and $\mathcal K(i,j)$ (right side of the summation). Therefore, the storage of all sets $\mathcal K'$ would require a size of $\mathcal O\left(\nMean{\widetilde{n}}^{d+1}\right)$. This is not efficient.
One loophole would be to store all possible indices $j$ in a hash table
$\mathcal D$:\cite{sedgewick_book}
\begin{lstlisting}
$a'_{:::} \leftarrow 0$
for $i$ in $\mathcal I$:
  for $l$ in $\mathcal J(i)$:
    for $k$ in $\mathcal K(i,j)$:
      for $j$ in $\mathcal D(i,k)$:
        $a'_{ilk} \leftarrow a'_{ilk} + h_{lj} a_{ijk}$
\end{lstlisting}
Using the hash table $\mathcal D$ reduces the memory requirements. However, the indices of the tensor $\vec a$ are now accessed in an arbitrary order which results in a very reduced performance. 

This problem is solved if the tensor transformation is performed over the \emph{last} dimension. Then, the tensor transformation reduces to \emph{one} matrix matrix product, because the multiindex $k$ does not exist:
\begin{lstlisting}
$a'_{::} \leftarrow 0$
for $i$ in $\mathcal I$:
  for $l$ in $\mathcal J(i)$:
    for $j$ in $\mathcal J(i)$:
        $a'_{il} \leftarrow a'_{il} + h_{lj} a_{ij}$
\end{lstlisting}
We assume that the pruned coefficient tensor $\vec a$ is properly sorted.
The storage of $\mathcal I$ and $\mathcal J(i)$ scales as $\mathcal O\left(\nMean{\widetilde{n}}^{d}\right)$ but it can be reduced by utilizing the fact that many indices in $\mathcal J(i)$ appear in ranges. Instead of storing all individual indices, we therefore store only the ranges. To give an example, $\mathcal J = \{1,2,3,4,6,7,8,10\}$ would be stored as $\{[1,4],[6,8],[10,10]\}$. 

We have shown that the transformation over the last dimension can be implemented in a very efficient way such that no auxiliary arrays of size larger than $\mathcal O\left(\nMean{\widetilde{n}}^{d}\right)$ need to be stored. In fact, the required storage of the needed arrays is much smaller due to the storage of the index ranges. Note that the sizes of the auxiliary arrays are never larger than the size of the \emph{pruned} basis. 

One can use this algorithm also for the transformation over other dimensions by simply permuting the tensor in a way that the dimension to be transformed over is the last one. After the transformation, the tensor is permuted back to the original order.
Finding the proper permutation is just a sorting operation and scales as $\mathcal O\left[\nMean{\widetilde{n}}^{d} \log\left(\nMean{\widetilde{n}}^d\right)\right]= \mathcal O\left[d \times  \nMean{\widetilde{n}}^d \log\left(\nMean{\widetilde{n}}\right)\right]$. This is much faster than the tensor transformation. Additionally, the permutations can be stored such that the sorting operations are needed only after a change of the basis. The memory requirements and the cost of the permutations scale as $\mathcal O(d\times \nMean{\widetilde{n}}^d)$. Compared to the storage requirements of the Lanczos-Arnoldi-propagator and the scaling of the tensor transformation, this is justifiable.
A permutation of the basis to obtain more efficient tensor transformations was already used in a  different application.\cite{pruned_prod_basis_mapping_carrington_2009}

To show the efficiency of the algorithm, we compare our approach to the unpruned algorithm that uses GEMM calls. For this, we do not prune the basis but store \emph{all} possible functions of the direct-product basis. We use \ce{NO2} (section \ref{sec:NO2}) as an example. The total size of the basis is $\sim 285\times 10^4$. The GEMM version (unpruned variant) needed $\unit[30.1]{s}$ for computing the matrix-vector product ($36$ sum terms) whereas the pruned variant takes $\unit[32.5]{s}$. The pruned variant is $\unit[8]\%$ slower. For a smaller basis with size $\sim 146\times 10^4$, the GEMM version needs $\unit[9.80]{s}$. The pruned variant needs $\unit[13.05]{s}$ and is $\unit[33]\%$ slower. For smaller vectors, more values fit into the cache of the CPU and the GEMM implementation is then more efficient.
Of course, in real simulations, the pruned variant will only be used for a pruned and not for a full basis and will therefore be much faster.

Note that this scheme for pruned tensor transformations is completely general; no assumptions about any structure in the pruning are made.
It could be further optimized by dividing the matrix into chunks that fit into the cache of the computer.\cite{gemm_goto_2008} 
Brown and Carrington have developed a different algorithm for a similar application, based on a more complicated recursive mapping strategy.\cite{symmetrized_gaussians_sop_carrington_2016} It remains to be seen which approach is faster.


\begin{thebibliography}{112}%
\makeatletter
\providecommand \@ifxundefined [1]{%
 \@ifx{#1\undefined}
}%
\providecommand \@ifnum [1]{%
 \ifnum #1\expandafter \@firstoftwo
 \else \expandafter \@secondoftwo
 \fi
}%
\providecommand \@ifx [1]{%
 \ifx #1\expandafter \@firstoftwo
 \else \expandafter \@secondoftwo
 \fi
}%
\providecommand \natexlab [1]{#1}%
\providecommand \enquote  [1]{``#1''}%
\providecommand \bibnamefont  [1]{#1}%
\providecommand \bibfnamefont [1]{#1}%
\providecommand \citenamefont [1]{#1}%
\providecommand \href@noop [0]{\@secondoftwo}%
\providecommand \href [0]{\begingroup \@sanitize@url \@href}%
\providecommand \@href[1]{\@@startlink{#1}\@@href}%
\providecommand \@@href[1]{\endgroup#1\@@endlink}%
\providecommand \@sanitize@url [0]{\catcode `\\12\catcode `\$12\catcode
  `\&12\catcode `\#12\catcode `\^12\catcode `\_12\catcode `\%12\relax}%
\providecommand \@@startlink[1]{}%
\providecommand \@@endlink[0]{}%
\providecommand \url  [0]{\begingroup\@sanitize@url \@url }%
\providecommand \@url [1]{\endgroup\@href {#1}{\urlprefix }}%
\providecommand \urlprefix  [0]{URL }%
\providecommand \Eprint [0]{\href }%
\providecommand \doibase [0]{http://dx.doi.org/}%
\providecommand \selectlanguage [0]{\@gobble}%
\providecommand \bibinfo  [0]{\@secondoftwo}%
\providecommand \bibfield  [0]{\@secondoftwo}%
\providecommand \translation [1]{[#1]}%
\providecommand \BibitemOpen [0]{}%
\providecommand \bibitemStop [0]{}%
\providecommand \bibitemNoStop [0]{.\EOS\space}%
\providecommand \EOS [0]{\spacefactor3000\relax}%
\providecommand \BibitemShut  [1]{\csname bibitem#1\endcsname}%
\let\auto@bib@innerbib\@empty
\bibitem [{\citenamefont {Tannor}(2007)}]{tannor_book}%
  \BibitemOpen
  \bibfield  {author} {\bibinfo {author} {\bibfnamefont {D.~J.}\ \bibnamefont
  {Tannor}},\ }\href@noop {} {\emph {\bibinfo {title} {Introduction to Quantum
  Mechanics: A Time-Dependent Perspective}}},\ \bibinfo {edition} {1st}\ ed.\
  (\bibinfo  {publisher} {University Science Books},\ \bibinfo {year}
  {2007})\BibitemShut {NoStop}%
\bibitem [{\citenamefont {Tizniti}\ \emph {et~al.}(2014)\citenamefont
  {Tizniti}, \citenamefont {Le~Picard}, \citenamefont {Lique}, \citenamefont
  {Berteloite}, \citenamefont {Canosa}, \citenamefont {Alexander},\ and\
  \citenamefont {Sims}}]{H2_F_low_temp_sims_2014}%
  \BibitemOpen
  \bibfield  {author} {\bibinfo {author} {\bibfnamefont {M.}~\bibnamefont
  {Tizniti}}, \bibinfo {author} {\bibfnamefont {S.~D.}\ \bibnamefont
  {Le~Picard}}, \bibinfo {author} {\bibfnamefont {F.}~\bibnamefont {Lique}},
  \bibinfo {author} {\bibfnamefont {C.}~\bibnamefont {Berteloite}}, \bibinfo
  {author} {\bibfnamefont {A.}~\bibnamefont {Canosa}}, \bibinfo {author}
  {\bibfnamefont {M.~H.}\ \bibnamefont {Alexander}}, \ and\ \bibinfo {author}
  {\bibfnamefont {I.~R.}\ \bibnamefont {Sims}},\ }\href {\doibase
  10.1038/nchem.1835} {\bibfield  {journal} {\bibinfo  {journal} {Nat. Chem.}\
  }\textbf {\bibinfo {volume} {6}},\ \bibinfo {pages} {141–145} (\bibinfo
  {year} {2014})}\BibitemShut {NoStop}%
\bibitem [{\citenamefont {Wu}, \citenamefont {Werner},\ and\ \citenamefont
  {Manthe}(2004)}]{H_CH4_manthe_2004}%
  \BibitemOpen
  \bibfield  {author} {\bibinfo {author} {\bibfnamefont {T.}~\bibnamefont
  {Wu}}, \bibinfo {author} {\bibfnamefont {H.-J.}\ \bibnamefont {Werner}}, \
  and\ \bibinfo {author} {\bibfnamefont {U.}~\bibnamefont {Manthe}},\ }\href
  {\doibase 10.1126/science.1104085} {\bibfield  {journal} {\bibinfo  {journal}
  {Science.}\ }\textbf {\bibinfo {volume} {306}},\ \bibinfo {pages}
  {2227–2229} (\bibinfo {year} {2004})}\BibitemShut {NoStop}%
\bibitem [{\citenamefont {Otto}\ \emph {et~al.}(2014)\citenamefont {Otto},
  \citenamefont {Ma}, \citenamefont {Ray}, \citenamefont {Daluz}, \citenamefont
  {Li}, \citenamefont {Guo},\ and\ \citenamefont
  {Continetti}}]{F_H2O_otto_2014}%
  \BibitemOpen
  \bibfield  {author} {\bibinfo {author} {\bibfnamefont {R.}~\bibnamefont
  {Otto}}, \bibinfo {author} {\bibfnamefont {J.}~\bibnamefont {Ma}}, \bibinfo
  {author} {\bibfnamefont {A.~W.}\ \bibnamefont {Ray}}, \bibinfo {author}
  {\bibfnamefont {J.~S.}\ \bibnamefont {Daluz}}, \bibinfo {author}
  {\bibfnamefont {J.}~\bibnamefont {Li}}, \bibinfo {author} {\bibfnamefont
  {H.}~\bibnamefont {Guo}}, \ and\ \bibinfo {author} {\bibfnamefont {R.~E.}\
  \bibnamefont {Continetti}},\ }\href {\doibase 10.1126/science.1247424}
  {\bibfield  {journal} {\bibinfo  {journal} {Science.}\ }\textbf {\bibinfo
  {volume} {343}},\ \bibinfo {pages} {396–399} (\bibinfo {year}
  {2014})}\BibitemShut {NoStop}%
\bibitem [{\citenamefont {Westermann}\ \emph {et~al.}(2014)\citenamefont
  {Westermann}, \citenamefont {Kim}, \citenamefont {Weichman}, \citenamefont
  {Hock}, \citenamefont {Yacovitch}, \citenamefont {Palma}, \citenamefont
  {Neumark},\ and\ \citenamefont {Manthe}}]{CH4F_manthe_2013}%
  \BibitemOpen
  \bibfield  {author} {\bibinfo {author} {\bibfnamefont {T.}~\bibnamefont
  {Westermann}}, \bibinfo {author} {\bibfnamefont {J.~B.}\ \bibnamefont {Kim}},
  \bibinfo {author} {\bibfnamefont {M.~L.}\ \bibnamefont {Weichman}}, \bibinfo
  {author} {\bibfnamefont {C.}~\bibnamefont {Hock}}, \bibinfo {author}
  {\bibfnamefont {T.~I.}\ \bibnamefont {Yacovitch}}, \bibinfo {author}
  {\bibfnamefont {J.}~\bibnamefont {Palma}}, \bibinfo {author} {\bibfnamefont
  {D.~M.}\ \bibnamefont {Neumark}}, \ and\ \bibinfo {author} {\bibfnamefont
  {U.}~\bibnamefont {Manthe}},\ }\href {\doibase 10.1002/anie.201307822}
  {\bibfield  {journal} {\bibinfo  {journal} {Angew. Chem. Int. Ed.}\ }\textbf
  {\bibinfo {volume} {53}},\ \bibinfo {pages} {1122–1126} (\bibinfo {year}
  {2014})}\BibitemShut {NoStop}%
\bibitem [{\citenamefont {Chen}\ \emph {et~al.}(2016)\citenamefont {Chen},
  \citenamefont {Shao}, \citenamefont {Chen}, \citenamefont {Yang},\ and\
  \citenamefont {Zhang}}]{C2H_H2_zhang_2016}%
  \BibitemOpen
  \bibfield  {author} {\bibinfo {author} {\bibfnamefont {L.}~\bibnamefont
  {Chen}}, \bibinfo {author} {\bibfnamefont {K.}~\bibnamefont {Shao}}, \bibinfo
  {author} {\bibfnamefont {J.}~\bibnamefont {Chen}}, \bibinfo {author}
  {\bibfnamefont {M.}~\bibnamefont {Yang}}, \ and\ \bibinfo {author}
  {\bibfnamefont {D.~H.}\ \bibnamefont {Zhang}},\ }\href {\doibase
  http://dx.doi.org/10.1063/1.4948996} {\bibfield  {journal} {\bibinfo
  {journal} {J. Chem. Phys.}\ }\textbf {\bibinfo {volume} {144}},\ \bibinfo
  {eid} {194309} (\bibinfo {year} {2016}),\
  http://dx.doi.org/10.1063/1.4948996}\BibitemShut {NoStop}%
\bibitem [{\citenamefont {Song}\ \emph {et~al.}(2014)\citenamefont {Song},
  \citenamefont {Li}, \citenamefont {Yang}, \citenamefont {Lu},\ and\
  \citenamefont {Guo}}]{H2_NH2_guo_2014}%
  \BibitemOpen
  \bibfield  {author} {\bibinfo {author} {\bibfnamefont {H.}~\bibnamefont
  {Song}}, \bibinfo {author} {\bibfnamefont {J.}~\bibnamefont {Li}}, \bibinfo
  {author} {\bibfnamefont {M.}~\bibnamefont {Yang}}, \bibinfo {author}
  {\bibfnamefont {Y.}~\bibnamefont {Lu}}, \ and\ \bibinfo {author}
  {\bibfnamefont {H.}~\bibnamefont {Guo}},\ }\href {\doibase
  10.1039/C4CP02227K} {\bibfield  {journal} {\bibinfo  {journal} {Phys. Chem.
  Chem. Phys.}\ }\textbf {\bibinfo {volume} {16}},\ \bibinfo {pages} {17770}
  (\bibinfo {year} {2014})}\BibitemShut {NoStop}%
\bibitem [{\citenamefont {Palma}\ and\ \citenamefont
  {Clary}(2000)}]{six_atom_dyn_reduced_dim_clary_2000}%
  \BibitemOpen
  \bibfield  {author} {\bibinfo {author} {\bibfnamefont {J.}~\bibnamefont
  {Palma}}\ and\ \bibinfo {author} {\bibfnamefont {D.~C.}\ \bibnamefont
  {Clary}},\ }\href {\doibase http://dx.doi.org/10.1063/1.480749} {\bibfield
  {journal} {\bibinfo  {journal} {J. Chem. Phys.}\ }\textbf {\bibinfo {volume}
  {112}},\ \bibinfo {pages} {1859} (\bibinfo {year} {2000})}\BibitemShut
  {NoStop}%
\bibitem [{\citenamefont {Zhang}\ \emph {et~al.}(2010)\citenamefont {Zhang},
  \citenamefont {Zhou}, \citenamefont {Wu}, \citenamefont {Lu}, \citenamefont
  {Pan}, \citenamefont {Fu}, \citenamefont {Shuai}, \citenamefont {Liu},
  \citenamefont {Liu}, \citenamefont {Zhang}, \citenamefont {Jiang},
  \citenamefont {Dai}, \citenamefont {Lee}, \citenamefont {Xie}, \citenamefont
  {Braams}, \citenamefont {Bowman}, \citenamefont {Collins}, \citenamefont
  {Zhang},\ and\ \citenamefont {Yang}}]{H_CD4_reduced_dim_zhang_2007}%
  \BibitemOpen
  \bibfield  {author} {\bibinfo {author} {\bibfnamefont {W.}~\bibnamefont
  {Zhang}}, \bibinfo {author} {\bibfnamefont {Y.}~\bibnamefont {Zhou}},
  \bibinfo {author} {\bibfnamefont {G.}~\bibnamefont {Wu}}, \bibinfo {author}
  {\bibfnamefont {Y.}~\bibnamefont {Lu}}, \bibinfo {author} {\bibfnamefont
  {H.}~\bibnamefont {Pan}}, \bibinfo {author} {\bibfnamefont {B.}~\bibnamefont
  {Fu}}, \bibinfo {author} {\bibfnamefont {Q.}~\bibnamefont {Shuai}}, \bibinfo
  {author} {\bibfnamefont {L.}~\bibnamefont {Liu}}, \bibinfo {author}
  {\bibfnamefont {S.}~\bibnamefont {Liu}}, \bibinfo {author} {\bibfnamefont
  {L.}~\bibnamefont {Zhang}}, \bibinfo {author} {\bibfnamefont
  {B.}~\bibnamefont {Jiang}}, \bibinfo {author} {\bibfnamefont
  {D.}~\bibnamefont {Dai}}, \bibinfo {author} {\bibfnamefont {S.-Y.}\
  \bibnamefont {Lee}}, \bibinfo {author} {\bibfnamefont {Z.}~\bibnamefont
  {Xie}}, \bibinfo {author} {\bibfnamefont {B.~J.}\ \bibnamefont {Braams}},
  \bibinfo {author} {\bibfnamefont {J.~M.}\ \bibnamefont {Bowman}}, \bibinfo
  {author} {\bibfnamefont {M.~A.}\ \bibnamefont {Collins}}, \bibinfo {author}
  {\bibfnamefont {D.~H.}\ \bibnamefont {Zhang}}, \ and\ \bibinfo {author}
  {\bibfnamefont {X.}~\bibnamefont {Yang}},\ }\href {\doibase
  10.1073/pnas.1006910107} {\bibfield  {journal} {\bibinfo  {journal} {Proc.
  Natl. Ac. Sci.}\ }\textbf {\bibinfo {volume} {107}},\ \bibinfo {pages}
  {12782} (\bibinfo {year} {2010})}\BibitemShut {NoStop}%
\bibitem [{\citenamefont {Baloïtcha}\ \emph {et~al.}(2002)\citenamefont
  {Baloïtcha}, \citenamefont {Lasorne}, \citenamefont {Lauvergnat},
  \citenamefont {Dive}, \citenamefont {Justum},\ and\ \citenamefont
  {Desouter-Lecomte}}]{red_dyn_h_transfer_descouter-lecomte_2002}%
  \BibitemOpen
  \bibfield  {author} {\bibinfo {author} {\bibfnamefont {E.}~\bibnamefont
  {Baloïtcha}}, \bibinfo {author} {\bibfnamefont {B.}~\bibnamefont {Lasorne}},
  \bibinfo {author} {\bibfnamefont {D.}~\bibnamefont {Lauvergnat}}, \bibinfo
  {author} {\bibfnamefont {G.}~\bibnamefont {Dive}}, \bibinfo {author}
  {\bibfnamefont {Y.}~\bibnamefont {Justum}}, \ and\ \bibinfo {author}
  {\bibfnamefont {M.}~\bibnamefont {Desouter-Lecomte}},\ }\href {\doibase
  http://dx.doi.org/10.1063/1.1481857} {\bibfield  {journal} {\bibinfo
  {journal} {J. Chem. Phys.}\ }\textbf {\bibinfo {volume} {117}},\ \bibinfo
  {pages} {727} (\bibinfo {year} {2002})}\BibitemShut {NoStop}%
\bibitem [{\citenamefont {Huarte-Larrañaga}\ and\ \citenamefont
  {Manthe}(2001)}]{CH4_H_full_versus_reduced_dim_qdyn_manthe_2001}%
  \BibitemOpen
  \bibfield  {author} {\bibinfo {author} {\bibfnamefont {F.}~\bibnamefont
  {Huarte-Larrañaga}}\ and\ \bibinfo {author} {\bibfnamefont {U.}~\bibnamefont
  {Manthe}},\ }\href {\doibase 10.1021/jp003579w} {\bibfield  {journal}
  {\bibinfo  {journal} {J. Phys. Chem. A}\ }\textbf {\bibinfo {volume} {105}},\
  \bibinfo {pages} {2522} (\bibinfo {year} {2001})},\ \Eprint
  {http://arxiv.org/abs/http://dx.doi.org/10.1021/jp003579w}
  {http://dx.doi.org/10.1021/jp003579w} \BibitemShut {NoStop}%
\bibitem [{\citenamefont {Vikár}, \citenamefont {Nagy},\ and\ \citenamefont
  {Lendvay}(2016)}]{CH4_H_reduced_dim_vs_full_dim_classical_lendvay_2016}%
  \BibitemOpen
  \bibfield  {author} {\bibinfo {author} {\bibfnamefont {A.}~\bibnamefont
  {Vikár}}, \bibinfo {author} {\bibfnamefont {T.}~\bibnamefont {Nagy}}, \ and\
  \bibinfo {author} {\bibfnamefont {G.}~\bibnamefont {Lendvay}},\ }\href
  {\doibase 10.1021/acs.jpca.6b00346} {\bibfield  {journal} {\bibinfo
  {journal} {J. Phys. Chem. A}\ }\textbf {\bibinfo {volume} {120}},\ \bibinfo
  {pages} {5083} (\bibinfo {year} {2016})}\BibitemShut {NoStop}%
\bibitem [{\citenamefont {Burghardt}, \citenamefont {Meyer},\ and\
  \citenamefont {Cederbaum}(1999)}]{G-MCTDH_Burghardt_1999}%
  \BibitemOpen
  \bibfield  {author} {\bibinfo {author} {\bibfnamefont {I.}~\bibnamefont
  {Burghardt}}, \bibinfo {author} {\bibfnamefont {H.-D.}\ \bibnamefont
  {Meyer}}, \ and\ \bibinfo {author} {\bibfnamefont {L.~S.}\ \bibnamefont
  {Cederbaum}},\ }\href {\doibase 10.1063/1.479574} {\bibfield  {journal}
  {\bibinfo  {journal} {J. Chem. Phys.}\ }\textbf {\bibinfo {volume} {111}},\
  \bibinfo {pages} {2927} (\bibinfo {year} {1999})}\BibitemShut {NoStop}%
\bibitem [{\citenamefont {Römer}, \citenamefont {Ruckenbauer},\ and\
  \citenamefont {Burghardt}(2013)}]{G-MCTDH_layer_Burghardt_2013}%
  \BibitemOpen
  \bibfield  {author} {\bibinfo {author} {\bibfnamefont {S.}~\bibnamefont
  {Römer}}, \bibinfo {author} {\bibfnamefont {M.}~\bibnamefont {Ruckenbauer}},
  \ and\ \bibinfo {author} {\bibfnamefont {I.}~\bibnamefont {Burghardt}},\
  }\href {\doibase 10.1063/1.4788830} {\bibfield  {journal} {\bibinfo
  {journal} {J. Chem. Phys.}\ }\textbf {\bibinfo {volume} {138}},\ \bibinfo
  {pages} {064106} (\bibinfo {year} {2013})}\BibitemShut {NoStop}%
\bibitem [{\citenamefont {Worth}, \citenamefont {Robb},\ and\ \citenamefont
  {Burghardt}(2004)}]{vMCG_Burghardt_2004}%
  \BibitemOpen
  \bibfield  {author} {\bibinfo {author} {\bibfnamefont {G.~A.}\ \bibnamefont
  {Worth}}, \bibinfo {author} {\bibfnamefont {M.~A.}\ \bibnamefont {Robb}}, \
  and\ \bibinfo {author} {\bibfnamefont {I.}~\bibnamefont {Burghardt}},\ }\href
  {\doibase 10.1039/b314253a} {\bibfield  {journal} {\bibinfo  {journal}
  {Faraday Discuss.}\ }\textbf {\bibinfo {volume} {127}},\ \bibinfo {pages}
  {307} (\bibinfo {year} {2004})}\BibitemShut {NoStop}%
\bibitem [{\citenamefont {Richings}\ \emph {et~al.}(2015)\citenamefont
  {Richings}, \citenamefont {Polyak}, \citenamefont {Spinlove}, \citenamefont
  {Worth}, \citenamefont {Burghardt},\ and\ \citenamefont
  {Lasorne}}]{vMCG_rev_lasorne_2015}%
  \BibitemOpen
  \bibfield  {author} {\bibinfo {author} {\bibfnamefont {G.}~\bibnamefont
  {Richings}}, \bibinfo {author} {\bibfnamefont {I.}~\bibnamefont {Polyak}},
  \bibinfo {author} {\bibfnamefont {K.}~\bibnamefont {Spinlove}}, \bibinfo
  {author} {\bibfnamefont {G.}~\bibnamefont {Worth}}, \bibinfo {author}
  {\bibfnamefont {I.}~\bibnamefont {Burghardt}}, \ and\ \bibinfo {author}
  {\bibfnamefont {B.}~\bibnamefont {Lasorne}},\ }\href {\doibase
  10.1080/0144235x.2015.1051354} {\bibfield  {journal} {\bibinfo  {journal}
  {Int. Rev. Phys. Chem.}\ }\textbf {\bibinfo {volume} {34}},\ \bibinfo {pages}
  {269–308} (\bibinfo {year} {2015})}\BibitemShut {NoStop}%
\bibitem [{\citenamefont {Wu}\ and\ \citenamefont
  {Batista}(2003)}]{matching_pursuit_SPO_batista_2003}%
  \BibitemOpen
  \bibfield  {author} {\bibinfo {author} {\bibfnamefont {Y.}~\bibnamefont
  {Wu}}\ and\ \bibinfo {author} {\bibfnamefont {V.~S.}\ \bibnamefont
  {Batista}},\ }\href {\doibase 10.1063/1.1560636} {\bibfield  {journal}
  {\bibinfo  {journal} {J. Chem. Phys.}\ }\textbf {\bibinfo {volume} {118}},\
  \bibinfo {pages} {6720} (\bibinfo {year} {2003})}\BibitemShut {NoStop}%
\bibitem [{\citenamefont {Martínez}, \citenamefont {Ben-Nun},\ and\
  \citenamefont {Levine}(1996)}]{aims_martinez_1996}%
  \BibitemOpen
  \bibfield  {author} {\bibinfo {author} {\bibfnamefont {T.~J.}\ \bibnamefont
  {Martínez}}, \bibinfo {author} {\bibfnamefont {M.}~\bibnamefont {Ben-Nun}},
  \ and\ \bibinfo {author} {\bibfnamefont {R.~D.}\ \bibnamefont {Levine}},\
  }\href {\doibase 10.1021/jp953105a} {\bibfield  {journal} {\bibinfo
  {journal} {J. Phys. Chem.}\ }\textbf {\bibinfo {volume} {100}},\ \bibinfo
  {pages} {7884} (\bibinfo {year} {1996})}\BibitemShut {NoStop}%
\bibitem [{\citenamefont {Ben-Nun}\ and\ \citenamefont
  {Martínez}(1998)}]{aims_validation_martinez_1998}%
  \BibitemOpen
  \bibfield  {author} {\bibinfo {author} {\bibfnamefont {M.}~\bibnamefont
  {Ben-Nun}}\ and\ \bibinfo {author} {\bibfnamefont {T.~J.}\ \bibnamefont
  {Martínez}},\ }\href {\doibase http://dx.doi.org/10.1063/1.476142}
  {\bibfield  {journal} {\bibinfo  {journal} {J. Chem. Phys.}\ }\textbf
  {\bibinfo {volume} {108}},\ \bibinfo {pages} {7244} (\bibinfo {year}
  {1998})}\BibitemShut {NoStop}%
\bibitem [{\citenamefont {Shalashilin}\ and\ \citenamefont
  {Child}(2004)}]{coupled_coherent_states_rev_shalashilin_2004}%
  \BibitemOpen
  \bibfield  {author} {\bibinfo {author} {\bibfnamefont {D.~V.}\ \bibnamefont
  {Shalashilin}}\ and\ \bibinfo {author} {\bibfnamefont {M.~S.}\ \bibnamefont
  {Child}},\ }\href {\doibase http://dx.doi.org/10.1016/j.chemphys.2004.06.013}
  {\bibfield  {journal} {\bibinfo  {journal} {Chem. Phys.}\ }\textbf {\bibinfo
  {volume} {304}},\ \bibinfo {pages} {103 } (\bibinfo {year}
  {2004})}\BibitemShut {NoStop}%
\bibitem [{\citenamefont {Meyer}, \citenamefont {Manthe},\ and\ \citenamefont
  {Cederbaum}(1990)}]{mctdh_cederbaum_1990}%
  \BibitemOpen
  \bibfield  {author} {\bibinfo {author} {\bibfnamefont {H.-D.}\ \bibnamefont
  {Meyer}}, \bibinfo {author} {\bibfnamefont {U.}~\bibnamefont {Manthe}}, \
  and\ \bibinfo {author} {\bibfnamefont {L.}~\bibnamefont {Cederbaum}},\ }\href
  {\doibase http://dx.doi.org/10.1016/0009-2614(90)87014-I} {\bibfield
  {journal} {\bibinfo  {journal} {Chem. Phys. Lett.}\ }\textbf {\bibinfo
  {volume} {165}},\ \bibinfo {pages} {73 } (\bibinfo {year}
  {1990})}\BibitemShut {NoStop}%
\bibitem [{\citenamefont {Beck}\ \emph {et~al.}(2000)\citenamefont {Beck},
  \citenamefont {J{\"a}ckle}, \citenamefont {Worth},\ and\ \citenamefont
  {Meyer}}]{mctdh_rev_meyer_2000}%
  \BibitemOpen
  \bibfield  {author} {\bibinfo {author} {\bibfnamefont {M.~H.}\ \bibnamefont
  {Beck}}, \bibinfo {author} {\bibfnamefont {A.}~\bibnamefont {J{\"a}ckle}},
  \bibinfo {author} {\bibfnamefont {G.~A.}\ \bibnamefont {Worth}}, \ and\
  \bibinfo {author} {\bibfnamefont {H.-D.}\ \bibnamefont {Meyer}},\ }\href@noop
  {} {\bibfield  {journal} {\bibinfo  {journal} {Phys. Rep.}\ }\textbf
  {\bibinfo {volume} {324}},\ \bibinfo {pages} {1} (\bibinfo {year}
  {2000})}\BibitemShut {NoStop}%
\bibitem [{\citenamefont {Wang}\ and\ \citenamefont
  {Thoss}(2003)}]{ml_mctdh_thoss_2003}%
  \BibitemOpen
  \bibfield  {author} {\bibinfo {author} {\bibfnamefont {H.}~\bibnamefont
  {Wang}}\ and\ \bibinfo {author} {\bibfnamefont {M.}~\bibnamefont {Thoss}},\
  }\href {\doibase http://dx.doi.org/10.1063/1.1580111} {\bibfield  {journal}
  {\bibinfo  {journal} {J. Chem. Phys.}\ }\textbf {\bibinfo {volume} {119}},\
  \bibinfo {pages} {1289} (\bibinfo {year} {2003})}\BibitemShut {NoStop}%
\bibitem [{\citenamefont {Manthe}(2008)}]{ml_mctdh_manthe_2008}%
  \BibitemOpen
  \bibfield  {author} {\bibinfo {author} {\bibfnamefont {U.}~\bibnamefont
  {Manthe}},\ }\href {\doibase http://dx.doi.org/10.1063/1.2902982} {\bibfield
  {journal} {\bibinfo  {journal} {J. Chem. Phys.}\ }\textbf {\bibinfo {volume}
  {128}},\ \bibinfo {eid} {164116} (\bibinfo {year} {2008}),\
  http://dx.doi.org/10.1063/1.2902982}\BibitemShut {NoStop}%
\bibitem [{\citenamefont {Vendrell}\ and\ \citenamefont
  {Meyer}(2011)}]{ml_mctdh_meyer_2011}%
  \BibitemOpen
  \bibfield  {author} {\bibinfo {author} {\bibfnamefont {O.}~\bibnamefont
  {Vendrell}}\ and\ \bibinfo {author} {\bibfnamefont {H.-D.}\ \bibnamefont
  {Meyer}},\ }\href {\doibase http://dx.doi.org/10.1063/1.3535541} {\bibfield
  {journal} {\bibinfo  {journal} {J. Chem. Phys.}\ }\textbf {\bibinfo {volume}
  {134}},\ \bibinfo {eid} {044135} (\bibinfo {year} {2011}),\
  http://dx.doi.org/10.1063/1.3535541}\BibitemShut {NoStop}%
\bibitem [{\citenamefont {Welsch}\ and\ \citenamefont
  {Manthe}(2012)}]{ml_mctdh_H_CH4_manthe_2012}%
  \BibitemOpen
  \bibfield  {author} {\bibinfo {author} {\bibfnamefont {R.}~\bibnamefont
  {Welsch}}\ and\ \bibinfo {author} {\bibfnamefont {U.}~\bibnamefont
  {Manthe}},\ }\href {\doibase http://dx.doi.org/10.1063/1.4772585} {\bibfield
  {journal} {\bibinfo  {journal} {J. Chem. Phys.}\ }\textbf {\bibinfo {volume}
  {137}},\ \bibinfo {eid} {244106} (\bibinfo {year} {2012}),\
  http://dx.doi.org/10.1063/1.4772585}\BibitemShut {NoStop}%
\bibitem [{\citenamefont {Meng}\ and\ \citenamefont
  {Meyer}(2014)}]{mlmctdh_versus_mctdh_H2COO_meyer_2014}%
  \BibitemOpen
  \bibfield  {author} {\bibinfo {author} {\bibfnamefont {Q.}~\bibnamefont
  {Meng}}\ and\ \bibinfo {author} {\bibfnamefont {H.-D.}\ \bibnamefont
  {Meyer}},\ }\href {\doibase 10.1063/1.4896201} {\bibfield  {journal}
  {\bibinfo  {journal} {J. Chem. Phys.}\ }\textbf {\bibinfo {volume} {141}},\
  \bibinfo {pages} {124309} (\bibinfo {year} {2014})}\BibitemShut {NoStop}%
\bibitem [{\citenamefont {Bačić}\ and\ \citenamefont
  {Light}(1986)}]{ray_dvr_light_1986}%
  \BibitemOpen
  \bibfield  {author} {\bibinfo {author} {\bibfnamefont {Z.}~\bibnamefont
  {Bačić}}\ and\ \bibinfo {author} {\bibfnamefont {J.~C.}\ \bibnamefont
  {Light}},\ }\href {\doibase http://dx.doi.org/10.1063/1.451824} {\bibfield
  {journal} {\bibinfo  {journal} {J. Chem. Phys.}\ }\textbf {\bibinfo {volume}
  {85}},\ \bibinfo {pages} {4594} (\bibinfo {year} {1986})}\BibitemShut
  {NoStop}%
\bibitem [{\citenamefont {Bačić}\ and\ \citenamefont
  {Light}(1987)}]{ray_dvr_appl_light_1987}%
  \BibitemOpen
  \bibfield  {author} {\bibinfo {author} {\bibfnamefont {Z.}~\bibnamefont
  {Bačić}}\ and\ \bibinfo {author} {\bibfnamefont {J.~C.}\ \bibnamefont
  {Light}},\ }\href {\doibase http://dx.doi.org/10.1063/1.452017} {\bibfield
  {journal} {\bibinfo  {journal} {J. Chem. Phys.}\ }\textbf {\bibinfo {volume}
  {86}},\ \bibinfo {pages} {3065} (\bibinfo {year} {1987})}\BibitemShut
  {NoStop}%
\bibitem [{\citenamefont {Bowman}\ and\ \citenamefont
  {Gazdy}(1991)}]{vib_spectra_contraction_gazdy_1990}%
  \BibitemOpen
  \bibfield  {author} {\bibinfo {author} {\bibfnamefont {J.~M.}\ \bibnamefont
  {Bowman}}\ and\ \bibinfo {author} {\bibfnamefont {B.}~\bibnamefont {Gazdy}},\
  }\href {\doibase http://dx.doi.org/10.1063/1.460361} {\bibfield  {journal}
  {\bibinfo  {journal} {J. Chem. Phys.}\ }\textbf {\bibinfo {volume} {94}},\
  \bibinfo {pages} {454} (\bibinfo {year} {1991})}\BibitemShut {NoStop}%
\bibitem [{\citenamefont {Bramley}\ and\ \citenamefont
  {Handy}(1993)}]{basis_contraction_rovib_handy_1992}%
  \BibitemOpen
  \bibfield  {author} {\bibinfo {author} {\bibfnamefont {M.~J.}\ \bibnamefont
  {Bramley}}\ and\ \bibinfo {author} {\bibfnamefont {N.~C.}\ \bibnamefont
  {Handy}},\ }\href {\doibase http://dx.doi.org/10.1063/1.464305} {\bibfield
  {journal} {\bibinfo  {journal} {J. Chem. Phys.}\ }\textbf {\bibinfo {volume}
  {98}},\ \bibinfo {pages} {1378} (\bibinfo {year} {1993})}\BibitemShut
  {NoStop}%
\bibitem [{\citenamefont {Viel}\ and\ \citenamefont
  {Leforestier}(2000)}]{contracted_basis_vibrational_spectrum_HFCO_leforestier_2000}%
  \BibitemOpen
  \bibfield  {author} {\bibinfo {author} {\bibfnamefont {A.}~\bibnamefont
  {Viel}}\ and\ \bibinfo {author} {\bibfnamefont {C.}~\bibnamefont
  {Leforestier}},\ }\href {\doibase http://dx.doi.org/10.1063/1.480674}
  {\bibfield  {journal} {\bibinfo  {journal} {J. Chem. Phys.}\ }\textbf
  {\bibinfo {volume} {112}},\ \bibinfo {pages} {1212} (\bibinfo {year}
  {2000})}\BibitemShut {NoStop}%
\bibitem [{\citenamefont
  {Luckhaus}(2000)}]{contraction_vibrational_generalized_coordinate_dvr_luckhaus_2000}%
  \BibitemOpen
  \bibfield  {author} {\bibinfo {author} {\bibfnamefont {D.}~\bibnamefont
  {Luckhaus}},\ }\href {\doibase http://dx.doi.org/10.1063/1.481924} {\bibfield
   {journal} {\bibinfo  {journal} {J. Chem. Phys.}\ }\textbf {\bibinfo {volume}
  {113}},\ \bibinfo {pages} {1329} (\bibinfo {year} {2000})}\BibitemShut
  {NoStop}%
\bibitem [{\citenamefont {Wang}\ and\ \citenamefont
  {Carrington}(2002)}]{contracted_bases_lanczos_carrington_2002}%
  \BibitemOpen
  \bibfield  {author} {\bibinfo {author} {\bibfnamefont {X.-G.}\ \bibnamefont
  {Wang}}\ and\ \bibinfo {author} {\bibfnamefont {T.}~\bibnamefont
  {Carrington}},\ }\href {\doibase 10.1063/1.1506911} {\bibfield  {journal}
  {\bibinfo  {journal} {J. Chem. Phys.}\ }\textbf {\bibinfo {volume} {117}},\
  \bibinfo {pages} {6923} (\bibinfo {year} {2002})}\BibitemShut {NoStop}%
\bibitem [{\citenamefont {Shimshovitz}(2015)}]{asaf_thesis}%
  \BibitemOpen
  \bibfield  {author} {\bibinfo {author} {\bibfnamefont {A.}~\bibnamefont
  {Shimshovitz}},\ }\emph {\bibinfo {title} {Phase Space Approach to Solving
  the Schrödinger Equation}},\ \href@noop {} {Ph.D. thesis},\ \bibinfo
  {school} {Weizmann Institute of Science} (\bibinfo {year} {2015})\BibitemShut
  {NoStop}%
\bibitem [{\citenamefont {Wang}\ and\ \citenamefont
  {Carrington}(2008)}]{CH5+_vib_carrington_2008}%
  \BibitemOpen
  \bibfield  {author} {\bibinfo {author} {\bibfnamefont {X.-G.}\ \bibnamefont
  {Wang}}\ and\ \bibinfo {author} {\bibfnamefont {T.}~\bibnamefont
  {Carrington}},\ }\href {\doibase 10.1063/1.3027825} {\bibfield  {journal}
  {\bibinfo  {journal} {J. Chem. Phys.}\ }\textbf {\bibinfo {volume} {129}},\
  \bibinfo {pages} {234102} (\bibinfo {year} {2008})}\BibitemShut {NoStop}%
\bibitem [{\citenamefont {Dawes}\ and\ \citenamefont
  {Carrington}(2005)}]{phase_space_localised_DVR_carrington_2005}%
  \BibitemOpen
  \bibfield  {author} {\bibinfo {author} {\bibfnamefont {R.}~\bibnamefont
  {Dawes}}\ and\ \bibinfo {author} {\bibfnamefont {T.}~\bibnamefont
  {Carrington}},\ }\href {\doibase 10.1063/1.1863935} {\bibfield  {journal}
  {\bibinfo  {journal} {J. Chem. Phys.}\ }\textbf {\bibinfo {volume} {122}},\
  \bibinfo {pages} {134101} (\bibinfo {year} {2005})}\BibitemShut {NoStop}%
\bibitem [{\citenamefont {Degani}, \citenamefont {Schiff},\ and\ \citenamefont
  {Tannor}(2005)}]{cubature_tannor_2005}%
  \BibitemOpen
  \bibfield  {author} {\bibinfo {author} {\bibfnamefont {I.}~\bibnamefont
  {Degani}}, \bibinfo {author} {\bibfnamefont {J.}~\bibnamefont {Schiff}}, \
  and\ \bibinfo {author} {\bibfnamefont {D.~J.}\ \bibnamefont {Tannor}},\
  }\href {\doibase 10.1007/s00211-005-0628-z} {\bibfield  {journal} {\bibinfo
  {journal} {Numer. Math.}\ }\textbf {\bibinfo {volume} {101}},\ \bibinfo
  {pages} {479} (\bibinfo {year} {2005})}\BibitemShut {NoStop}%
\bibitem [{\citenamefont {Degani}\ and\ \citenamefont
  {Tannor}(2006)}]{multidimensional_dvr_cubature_tannor_2006}%
  \BibitemOpen
  \bibfield  {author} {\bibinfo {author} {\bibfnamefont {I.}~\bibnamefont
  {Degani}}\ and\ \bibinfo {author} {\bibfnamefont {D.~J.}\ \bibnamefont
  {Tannor}},\ }\href {\doibase 10.1021/jp056587r} {\bibfield  {journal}
  {\bibinfo  {journal} {J. Phys. Chem. A}\ }\textbf {\bibinfo {volume} {110}},\
  \bibinfo {pages} {5395} (\bibinfo {year} {2006})}\BibitemShut {NoStop}%
\bibitem [{\citenamefont {Maynard}, \citenamefont {Wyatt},\ and\ \citenamefont
  {Iung}(1997)}]{eigenstates_fluoroform_efficient_vib_basis_lung_1997}%
  \BibitemOpen
  \bibfield  {author} {\bibinfo {author} {\bibfnamefont {A.}~\bibnamefont
  {Maynard}}, \bibinfo {author} {\bibfnamefont {R.~E.}\ \bibnamefont {Wyatt}},
  \ and\ \bibinfo {author} {\bibfnamefont {C.}~\bibnamefont {Iung}},\ }\href
  {\doibase http://dx.doi.org/10.1063/1.473850} {\bibfield  {journal} {\bibinfo
   {journal} {J. Chem. Phys.}\ }\textbf {\bibinfo {volume} {106}},\ \bibinfo
  {pages} {9483} (\bibinfo {year} {1997})}\BibitemShut {NoStop}%
\bibitem [{\citenamefont {Bégué}\ \emph {et~al.}(2007)\citenamefont
  {Bégué}, \citenamefont {Gohaud}, \citenamefont {Pouchan}, \citenamefont
  {Cassam-Chenaï},\ and\ \citenamefont
  {Liévin}}]{vibrational_selected_ci_vmwci_lievin_2007}%
  \BibitemOpen
  \bibfield  {author} {\bibinfo {author} {\bibfnamefont {D.}~\bibnamefont
  {Bégué}}, \bibinfo {author} {\bibfnamefont {N.}~\bibnamefont {Gohaud}},
  \bibinfo {author} {\bibfnamefont {C.}~\bibnamefont {Pouchan}}, \bibinfo
  {author} {\bibfnamefont {P.}~\bibnamefont {Cassam-Chenaï}}, \ and\ \bibinfo
  {author} {\bibfnamefont {J.}~\bibnamefont {Liévin}},\ }\href {\doibase
  http://dx.doi.org/10.1063/1.2795711} {\bibfield  {journal} {\bibinfo
  {journal} {J. Chem. Phys.}\ }\textbf {\bibinfo {volume} {127}},\ \bibinfo
  {eid} {164115} (\bibinfo {year} {2007}),\
  http://dx.doi.org/10.1063/1.2795711}\BibitemShut {NoStop}%
\bibitem [{\citenamefont {Halverson}\ and\ \citenamefont
  {Poirier}(2015{\natexlab{a}})}]{benzene_hybrid_truncation_scheme_HO_basis_poirier_2015}%
  \BibitemOpen
  \bibfield  {author} {\bibinfo {author} {\bibfnamefont {T.}~\bibnamefont
  {Halverson}}\ and\ \bibinfo {author} {\bibfnamefont {B.}~\bibnamefont
  {Poirier}},\ }\href {\doibase 10.1021/acs.jpca.5b07868} {\bibfield  {journal}
  {\bibinfo  {journal} {J. Phys. Chem. A}\ }\textbf {\bibinfo {volume} {119}},\
  \bibinfo {pages} {12417–12433} (\bibinfo {year}
  {2015}{\natexlab{a}})}\BibitemShut {NoStop}%
\bibitem [{\citenamefont {Shimshovitz}, \citenamefont {Bačić},\ and\
  \citenamefont {Tannor}(2014)}]{pvb_LiCN_tannor_2014}%
  \BibitemOpen
  \bibfield  {author} {\bibinfo {author} {\bibfnamefont {A.}~\bibnamefont
  {Shimshovitz}}, \bibinfo {author} {\bibfnamefont {Z.}~\bibnamefont
  {Bačić}}, \ and\ \bibinfo {author} {\bibfnamefont {D.~J.}\ \bibnamefont
  {Tannor}},\ }\href {\doibase 10.1063/1.4902553} {\bibfield  {journal}
  {\bibinfo  {journal} {J. Chem. Phys.}\ }\textbf {\bibinfo {volume} {141}},\
  \bibinfo {pages} {234106} (\bibinfo {year} {2014})}\BibitemShut {NoStop}%
\bibitem [{\citenamefont {Dawes}\ and\ \citenamefont
  {Carrington}(2004)}]{multidim_dvr_simultaneous_diagonalisation_carrington_2004}%
  \BibitemOpen
  \bibfield  {author} {\bibinfo {author} {\bibfnamefont {R.}~\bibnamefont
  {Dawes}}\ and\ \bibinfo {author} {\bibfnamefont {T.}~\bibnamefont
  {Carrington}},\ }\href {\doibase 10.1063/1.1758941} {\bibfield  {journal}
  {\bibinfo  {journal} {J. Chem. Phys.}\ }\textbf {\bibinfo {volume} {121}},\
  \bibinfo {pages} {726} (\bibinfo {year} {2004})}\BibitemShut {NoStop}%
\bibitem [{\citenamefont {Avila}\ and\ \citenamefont
  {Carrington}(2009)}]{nonprod_grid_carrington_2009}%
  \BibitemOpen
  \bibfield  {author} {\bibinfo {author} {\bibfnamefont {G.}~\bibnamefont
  {Avila}}\ and\ \bibinfo {author} {\bibfnamefont {T.}~\bibnamefont
  {Carrington}},\ }\href {\doibase 10.1063/1.3246593} {\bibfield  {journal}
  {\bibinfo  {journal} {J. Chem. Phys.}\ }\textbf {\bibinfo {volume} {131}},\
  \bibinfo {pages} {174103} (\bibinfo {year} {2009})}\BibitemShut {NoStop}%
\bibitem [{\citenamefont {Lauvergnat}\ and\ \citenamefont
  {Nauts}(2014)}]{sparse_grid_smolyak_nauts_2014}%
  \BibitemOpen
  \bibfield  {author} {\bibinfo {author} {\bibfnamefont {D.}~\bibnamefont
  {Lauvergnat}}\ and\ \bibinfo {author} {\bibfnamefont {A.}~\bibnamefont
  {Nauts}},\ }\href {\doibase http://dx.doi.org/10.1016/j.saa.2013.05.068}
  {\bibfield  {journal} {\bibinfo  {journal} {Spectrochim. Acta, Part A}\
  }\textbf {\bibinfo {volume} {119}},\ \bibinfo {pages} {18 } (\bibinfo {year}
  {2014})}\BibitemShut {NoStop}%
\bibitem [{\citenamefont {Colbert}\ and\ \citenamefont
  {Miller}(1992)}]{sinc_dvr_miller_1992}%
  \BibitemOpen
  \bibfield  {author} {\bibinfo {author} {\bibfnamefont {D.~T.}\ \bibnamefont
  {Colbert}}\ and\ \bibinfo {author} {\bibfnamefont {W.~H.}\ \bibnamefont
  {Miller}},\ }\href {\doibase 10.1063/1.462100} {\bibfield  {journal}
  {\bibinfo  {journal} {J. Chem. Phys.}\ }\textbf {\bibinfo {volume} {96}},\
  \bibinfo {pages} {1982} (\bibinfo {year} {1992})}\BibitemShut {NoStop}%
\bibitem [{\citenamefont {McLeod}\ and\ \citenamefont
  {Carrington}(2010)}]{pruning_tdse_carrington_2010}%
  \BibitemOpen
  \bibfield  {author} {\bibinfo {author} {\bibfnamefont {M.}~\bibnamefont
  {McLeod}}\ and\ \bibinfo {author} {\bibfnamefont {T.}~\bibnamefont
  {Carrington}},\ }\href {\doibase 10.1016/j.cplett.2010.10.034} {\bibfield
  {journal} {\bibinfo  {journal} {Chem. Phys. Lett.}\ }\textbf {\bibinfo
  {volume} {501}},\ \bibinfo {pages} {130–133} (\bibinfo {year}
  {2010})}\BibitemShut {NoStop}%
\bibitem [{\citenamefont {Gradinaru}(2007)}]{sparse_grid_fft_gradinaru_2007}%
  \BibitemOpen
  \bibfield  {author} {\bibinfo {author} {\bibfnamefont {V.}~\bibnamefont
  {Gradinaru}},\ }\href {\doibase 10.1007/s00607-007-0225-3} {\bibfield
  {journal} {\bibinfo  {journal} {Computing.}\ }\textbf {\bibinfo {volume}
  {80}},\ \bibinfo {pages} {1} (\bibinfo {year} {2007})}\BibitemShut {NoStop}%
\bibitem [{\citenamefont {Zhang}\ and\ \citenamefont
  {Zhang}(1994)}]{H2_OH_diatom_diatom_reaction_treatment_zhang_1994}%
  \BibitemOpen
  \bibfield  {author} {\bibinfo {author} {\bibfnamefont {D.~H.}\ \bibnamefont
  {Zhang}}\ and\ \bibinfo {author} {\bibfnamefont {J.~Z.~H.}\ \bibnamefont
  {Zhang}},\ }\href {\doibase 10.1063/1.467808} {\bibfield  {journal} {\bibinfo
   {journal} {J. Chem. Phys.}\ }\textbf {\bibinfo {volume} {101}},\ \bibinfo
  {pages} {1146} (\bibinfo {year} {1994})}\BibitemShut {NoStop}%
\bibitem [{\citenamefont {Hartke}(2006)}]{proDG_hartke_2006}%
  \BibitemOpen
  \bibfield  {author} {\bibinfo {author} {\bibfnamefont {B.}~\bibnamefont
  {Hartke}},\ }\href {\doibase 10.1039/b606376d} {\bibfield  {journal}
  {\bibinfo  {journal} {Phys. Chem. Chem. Phys.}\ }\textbf {\bibinfo {volume}
  {8}},\ \bibinfo {pages} {3627} (\bibinfo {year} {2006})}\BibitemShut
  {NoStop}%
\bibitem [{\citenamefont {Sielk}\ \emph {et~al.}(2009)\citenamefont {Sielk},
  \citenamefont {von Horsten}, \citenamefont {Krüger}, \citenamefont
  {Schneider},\ and\ \citenamefont {Hartke}}]{proDG_hartke_2008}%
  \BibitemOpen
  \bibfield  {author} {\bibinfo {author} {\bibfnamefont {J.}~\bibnamefont
  {Sielk}}, \bibinfo {author} {\bibfnamefont {H.~F.}\ \bibnamefont {von
  Horsten}}, \bibinfo {author} {\bibfnamefont {F.}~\bibnamefont {Krüger}},
  \bibinfo {author} {\bibfnamefont {R.}~\bibnamefont {Schneider}}, \ and\
  \bibinfo {author} {\bibfnamefont {B.}~\bibnamefont {Hartke}},\ }\href
  {\doibase 10.1039/b814315c} {\bibfield  {journal} {\bibinfo  {journal} {Phys.
  Chem. Chem. Phys.}\ }\textbf {\bibinfo {volume} {11}},\ \bibinfo {pages}
  {463–475} (\bibinfo {year} {2009})}\BibitemShut {NoStop}%
\bibitem [{\citenamefont {McCormack}(2006)}]{pruning_mccormack_2006}%
  \BibitemOpen
  \bibfield  {author} {\bibinfo {author} {\bibfnamefont {D.~A.}\ \bibnamefont
  {McCormack}},\ }\href {\doibase http://dx.doi.org/10.1063/1.2196889}
  {\bibfield  {journal} {\bibinfo  {journal} {J. Chem. Phys.}\ }\textbf
  {\bibinfo {volume} {124}},\ \bibinfo {eid} {204101} (\bibinfo {year}
  {2006}),\ http://dx.doi.org/10.1063/1.2196889}\BibitemShut {NoStop}%
\bibitem [{\citenamefont {Pettey}\ and\ \citenamefont
  {Wyatt}(2006)}]{pruning_wyatt_2006}%
  \BibitemOpen
  \bibfield  {author} {\bibinfo {author} {\bibfnamefont {L.~R.}\ \bibnamefont
  {Pettey}}\ and\ \bibinfo {author} {\bibfnamefont {R.~E.}\ \bibnamefont
  {Wyatt}},\ }\href {\doibase http://dx.doi.org/10.1016/j.cplett.2006.04.081}
  {\bibfield  {journal} {\bibinfo  {journal} {Chem. Phys. Lett.}\ }\textbf
  {\bibinfo {volume} {424}},\ \bibinfo {pages} {443 } (\bibinfo {year}
  {2006})}\BibitemShut {NoStop}%
\bibitem [{\citenamefont {Pettey}\ and\ \citenamefont
  {Wyatt}(2007)}]{pruning_wyatt_2007}%
  \BibitemOpen
  \bibfield  {author} {\bibinfo {author} {\bibfnamefont {L.~R.}\ \bibnamefont
  {Pettey}}\ and\ \bibinfo {author} {\bibfnamefont {R.~E.}\ \bibnamefont
  {Wyatt}},\ }\href {\doibase 10.1002/qua.21301} {\bibfield  {journal}
  {\bibinfo  {journal} {Int. J. Quantum Chem.}\ }\textbf {\bibinfo {volume}
  {107}},\ \bibinfo {pages} {1566} (\bibinfo {year} {2007})}\BibitemShut
  {NoStop}%
\bibitem [{\citenamefont {Davis}\ and\ \citenamefont
  {Heller}(1979)}]{semicl_gauss_heller_1979}%
  \BibitemOpen
  \bibfield  {author} {\bibinfo {author} {\bibfnamefont {M.~J.}\ \bibnamefont
  {Davis}}\ and\ \bibinfo {author} {\bibfnamefont {E.~J.}\ \bibnamefont
  {Heller}},\ }\href {\doibase 10.1063/1.438727} {\bibfield  {journal}
  {\bibinfo  {journal} {J. Chem. Phys.}\ }\textbf {\bibinfo {volume} {71}},\
  \bibinfo {pages} {3383} (\bibinfo {year} {1979})}\BibitemShut {NoStop}%
\bibitem [{\citenamefont {Poirier}(2003)}]{weylet_1_poirier_2003}%
  \BibitemOpen
  \bibfield  {author} {\bibinfo {author} {\bibfnamefont {B.}~\bibnamefont
  {Poirier}},\ }\href {\doibase 10.1142/S0219633603000380} {\bibfield
  {journal} {\bibinfo  {journal} {J. Theor. Comp. Chem.}\ }\textbf {\bibinfo
  {volume} {02}},\ \bibinfo {pages} {65} (\bibinfo {year} {2003})}\BibitemShut
  {NoStop}%
\bibitem [{\citenamefont {Poirier}\ and\ \citenamefont
  {Salam}(2004{\natexlab{a}})}]{weylet_2_poirier_2004}%
  \BibitemOpen
  \bibfield  {author} {\bibinfo {author} {\bibfnamefont {B.}~\bibnamefont
  {Poirier}}\ and\ \bibinfo {author} {\bibfnamefont {A.}~\bibnamefont
  {Salam}},\ }\href {\doibase 10.1063/1.1767511} {\bibfield  {journal}
  {\bibinfo  {journal} {J. Chem. Phys.}\ }\textbf {\bibinfo {volume} {121}},\
  \bibinfo {pages} {1690} (\bibinfo {year} {2004}{\natexlab{a}})}\BibitemShut
  {NoStop}%
\bibitem [{\citenamefont {Poirier}\ and\ \citenamefont
  {Salam}(2004{\natexlab{b}})}]{weylet_3_poirier_2004}%
  \BibitemOpen
  \bibfield  {author} {\bibinfo {author} {\bibfnamefont {B.}~\bibnamefont
  {Poirier}}\ and\ \bibinfo {author} {\bibfnamefont {A.}~\bibnamefont
  {Salam}},\ }\href {\doibase 10.1063/1.1767512} {\bibfield  {journal}
  {\bibinfo  {journal} {J. Chem. Phys.}\ }\textbf {\bibinfo {volume} {121}},\
  \bibinfo {pages} {1704} (\bibinfo {year} {2004}{\natexlab{b}})}\BibitemShut
  {NoStop}%
\bibitem [{\citenamefont {Shimshovitz}\ and\ \citenamefont
  {Tannor}(2012{\natexlab{a}})}]{pvb_tannor_2012}%
  \BibitemOpen
  \bibfield  {author} {\bibinfo {author} {\bibfnamefont {A.}~\bibnamefont
  {Shimshovitz}}\ and\ \bibinfo {author} {\bibfnamefont {D.~J.}\ \bibnamefont
  {Tannor}},\ }\href {\doibase 10.1103/physrevlett.109.070402} {\bibfield
  {journal} {\bibinfo  {journal} {Phys. Rev. Lett.}\ }\textbf {\bibinfo
  {volume} {109}} (\bibinfo {year} {2012}{\natexlab{a}}),\
  10.1103/physrevlett.109.070402}\BibitemShut {NoStop}%
\bibitem [{\citenamefont {Lombardini}\ and\ \citenamefont
  {Poirier}(2006)}]{weylets_Ne2_poirier_2006}%
  \BibitemOpen
  \bibfield  {author} {\bibinfo {author} {\bibfnamefont {R.}~\bibnamefont
  {Lombardini}}\ and\ \bibinfo {author} {\bibfnamefont {B.}~\bibnamefont
  {Poirier}},\ }\href {\doibase 10.1063/1.2187473} {\bibfield  {journal}
  {\bibinfo  {journal} {J. Chem. Phys.}\ }\textbf {\bibinfo {volume} {124}},\
  \bibinfo {pages} {144107} (\bibinfo {year} {2006})}\BibitemShut {NoStop}%
\bibitem [{\citenamefont {Halverson}\ and\ \citenamefont
  {Poirier}(2012)}]{symmetr_gauss_weylet_poirier_2012}%
  \BibitemOpen
  \bibfield  {author} {\bibinfo {author} {\bibfnamefont {T.}~\bibnamefont
  {Halverson}}\ and\ \bibinfo {author} {\bibfnamefont {B.}~\bibnamefont
  {Poirier}},\ }\href {\doibase 10.1063/1.4769402} {\bibfield  {journal}
  {\bibinfo  {journal} {J. Chem. Phys.}\ }\textbf {\bibinfo {volume} {137}},\
  \bibinfo {pages} {224101} (\bibinfo {year} {2012})}\BibitemShut {NoStop}%
\bibitem [{\citenamefont {Halverson}\ and\ \citenamefont
  {Poirier}(2014)}]{symmetr_gauss_appl_CH2NH_poirier_2014}%
  \BibitemOpen
  \bibfield  {author} {\bibinfo {author} {\bibfnamefont {T.}~\bibnamefont
  {Halverson}}\ and\ \bibinfo {author} {\bibfnamefont {B.}~\bibnamefont
  {Poirier}},\ }\href {\doibase 10.1063/1.4879216} {\bibfield  {journal}
  {\bibinfo  {journal} {J. Chem. Phys.}\ }\textbf {\bibinfo {volume} {140}},\
  \bibinfo {pages} {204112} (\bibinfo {year} {2014})}\BibitemShut {NoStop}%
\bibitem [{\citenamefont {Halverson}\ and\ \citenamefont
  {Poirier}(2015{\natexlab{b}})}]{symmetrized_gaussians_acetonitrile_poirier_2015}%
  \BibitemOpen
  \bibfield  {author} {\bibinfo {author} {\bibfnamefont {T.}~\bibnamefont
  {Halverson}}\ and\ \bibinfo {author} {\bibfnamefont {B.}~\bibnamefont
  {Poirier}},\ }\href {\doibase http://dx.doi.org/10.1016/j.cplett.2015.02.004}
  {\bibfield  {journal} {\bibinfo  {journal} {Chem. Phys. Lett.}\ }\textbf
  {\bibinfo {volume} {624}},\ \bibinfo {pages} {37 } (\bibinfo {year}
  {2015}{\natexlab{b}})}\BibitemShut {NoStop}%
\bibitem [{\citenamefont {Takemoto}, \citenamefont {Shimshovitz},\ and\
  \citenamefont {Tannor}(2012)}]{pvb_edyn_takemoto_tannor_2012}%
  \BibitemOpen
  \bibfield  {author} {\bibinfo {author} {\bibfnamefont {N.}~\bibnamefont
  {Takemoto}}, \bibinfo {author} {\bibfnamefont {A.}~\bibnamefont
  {Shimshovitz}}, \ and\ \bibinfo {author} {\bibfnamefont {D.~J.}\ \bibnamefont
  {Tannor}},\ }\href {\doibase 10.1063/1.4732306} {\bibfield  {journal}
  {\bibinfo  {journal} {J. Chem. Phys.}\ }\textbf {\bibinfo {volume} {137}},\
  \bibinfo {pages} {011102} (\bibinfo {year} {2012})}\BibitemShut {NoStop}%
\bibitem [{\citenamefont {Assémat}, \citenamefont {Machnes},\ and\
  \citenamefont {Tannor}(2015)}]{pvb_edyn_tannor_2015}%
  \BibitemOpen
  \bibfield  {author} {\bibinfo {author} {\bibfnamefont {E.}~\bibnamefont
  {Assémat}}, \bibinfo {author} {\bibfnamefont {S.}~\bibnamefont {Machnes}}, \
  and\ \bibinfo {author} {\bibfnamefont {D.}~\bibnamefont {Tannor}},\
  }\href@noop {} {\enquote {\bibinfo {title} {Double ionization of {Helium}
  from a phase space perspective},}\ } (\bibinfo {year} {2015}),\ \Eprint
  {http://arxiv.org/abs/arXiv:1502.05165} {arXiv:1502.05165} \BibitemShut
  {NoStop}%
\bibitem [{\citenamefont {Brown}\ and\ \citenamefont
  {Carrington~Jr.}(2016)}]{symmetrized_gaussians_sop_carrington_2016}%
  \BibitemOpen
  \bibfield  {author} {\bibinfo {author} {\bibfnamefont {J.}~\bibnamefont
  {Brown}}\ and\ \bibinfo {author} {\bibfnamefont {T.}~\bibnamefont
  {Carrington~Jr.}},\ }\href {\doibase http://dx.doi.org/10.1063/1.4954721}
  {\enquote {\bibinfo {title} {Assessing the utility of phase-space-localized
  basis functions: Exploiting direct product structure and a new basis function
  selection procedure},}\ } (\bibinfo {year} {2016})\BibitemShut {NoStop}%
\bibitem [{\citenamefont {Christiansen}(2012)}]{vcc_rev_christiansen_2012}%
  \BibitemOpen
  \bibfield  {author} {\bibinfo {author} {\bibfnamefont {O.}~\bibnamefont
  {Christiansen}},\ }\href {\doibase 10.1039/c2cp40090a} {\bibfield  {journal}
  {\bibinfo  {journal} {Phys. Chem. Chem. Phys.}\ }\textbf {\bibinfo {volume}
  {14}},\ \bibinfo {pages} {6672} (\bibinfo {year} {2012})}\BibitemShut
  {NoStop}%
\bibitem [{\citenamefont {Leclerc}\ and\ \citenamefont
  {Carrington}(2014)}]{cp_decomposition_carrington_2014}%
  \BibitemOpen
  \bibfield  {author} {\bibinfo {author} {\bibfnamefont {A.}~\bibnamefont
  {Leclerc}}\ and\ \bibinfo {author} {\bibfnamefont {T.}~\bibnamefont
  {Carrington}},\ }\href {\doibase 10.1063/1.4871981} {\bibfield  {journal}
  {\bibinfo  {journal} {J. Chem. Phys.}\ }\textbf {\bibinfo {volume} {140}},\
  \bibinfo {pages} {174111} (\bibinfo {year} {2014})}\BibitemShut {NoStop}%
\bibitem [{\citenamefont {Jäckle}\ and\ \citenamefont
  {Meyer}(1996)}]{potfit_meyer_1996}%
  \BibitemOpen
  \bibfield  {author} {\bibinfo {author} {\bibfnamefont {A.}~\bibnamefont
  {Jäckle}}\ and\ \bibinfo {author} {\bibfnamefont {H.}~\bibnamefont
  {Meyer}},\ }\href {\doibase http://dx.doi.org/10.1063/1.471513} {\bibfield
  {journal} {\bibinfo  {journal} {J. Chem. Phys.}\ }\textbf {\bibinfo {volume}
  {104}},\ \bibinfo {pages} {7974} (\bibinfo {year} {1996})}\BibitemShut
  {NoStop}%
\bibitem [{\citenamefont {Peláez}\ and\ \citenamefont
  {Meyer}(2013)}]{multigrid_potfit_meyer_2013}%
  \BibitemOpen
  \bibfield  {author} {\bibinfo {author} {\bibfnamefont {D.}~\bibnamefont
  {Peláez}}\ and\ \bibinfo {author} {\bibfnamefont {H.-D.}\ \bibnamefont
  {Meyer}},\ }\href {\doibase 10.1063/1.4773021} {\bibfield  {journal}
  {\bibinfo  {journal} {J. Chem. Phys.}\ }\textbf {\bibinfo {volume} {138}},\
  \bibinfo {pages} {014108} (\bibinfo {year} {2013})}\BibitemShut {NoStop}%
\bibitem [{\citenamefont {Otto}(2014)}]{mctdh_multilayer_potfit_otto_2014}%
  \BibitemOpen
  \bibfield  {author} {\bibinfo {author} {\bibfnamefont {F.}~\bibnamefont
  {Otto}},\ }\href {\doibase 10.1063/1.4856135} {\bibfield  {journal} {\bibinfo
   {journal} {J. Chem. Phys.}\ }\textbf {\bibinfo {volume} {140}},\ \bibinfo
  {pages} {014106} (\bibinfo {year} {2014})}\BibitemShut {NoStop}%
\bibitem [{\citenamefont {Manzhos}\ and\ \citenamefont
  {Carrington}(2006)}]{pes_neural_network_sum_of_products_carrington_2006}%
  \BibitemOpen
  \bibfield  {author} {\bibinfo {author} {\bibfnamefont {S.}~\bibnamefont
  {Manzhos}}\ and\ \bibinfo {author} {\bibfnamefont {T.}~\bibnamefont
  {Carrington}},\ }\href {\doibase 10.1063/1.2387950} {\bibfield  {journal}
  {\bibinfo  {journal} {J. Chem. Phys.}\ }\textbf {\bibinfo {volume} {125}},\
  \bibinfo {pages} {194105} (\bibinfo {year} {2006})}\BibitemShut {NoStop}%
\bibitem [{\citenamefont {Koch}\ and\ \citenamefont
  {Zhang}(2014)}]{sop_pes_neural_networks_zhang_2014}%
  \BibitemOpen
  \bibfield  {author} {\bibinfo {author} {\bibfnamefont {W.}~\bibnamefont
  {Koch}}\ and\ \bibinfo {author} {\bibfnamefont {D.~H.}\ \bibnamefont
  {Zhang}},\ }\href {\doibase http://dx.doi.org/10.1063/1.4887508} {\bibfield
  {journal} {\bibinfo  {journal} {J. Chem. Phys.}\ }\textbf {\bibinfo {volume}
  {141}},\ \bibinfo {eid} {021101} (\bibinfo {year} {2014}),\
  http://dx.doi.org/10.1063/1.4887508}\BibitemShut {NoStop}%
\bibitem [{\citenamefont {Avila}\ and\ \citenamefont
  {Carrington}(2015)}]{pes_sum_of_products_smolyak_carrington_2015}%
  \BibitemOpen
  \bibfield  {author} {\bibinfo {author} {\bibfnamefont {G.}~\bibnamefont
  {Avila}}\ and\ \bibinfo {author} {\bibfnamefont {T.}~\bibnamefont
  {Carrington}},\ }\href {\doibase 10.1063/1.4926651} {\bibfield  {journal}
  {\bibinfo  {journal} {J. Chem. Phys.}\ }\textbf {\bibinfo {volume} {143}},\
  \bibinfo {pages} {044106} (\bibinfo {year} {2015})}\BibitemShut {NoStop}%
\bibitem [{\citenamefont {Ziegler}\ and\ \citenamefont
  {Rauhut}(2016)}]{pes_sop_from_multimode_fit_rauhut_2016}%
  \BibitemOpen
  \bibfield  {author} {\bibinfo {author} {\bibfnamefont {B.}~\bibnamefont
  {Ziegler}}\ and\ \bibinfo {author} {\bibfnamefont {G.}~\bibnamefont
  {Rauhut}},\ }\href {\doibase http://dx.doi.org/10.1063/1.4943985} {\bibfield
  {journal} {\bibinfo  {journal} {J. Chem. Phys.}\ }\textbf {\bibinfo {volume}
  {144}},\ \bibinfo {eid} {114114} (\bibinfo {year} {2016}),\
  http://dx.doi.org/10.1063/1.4943985}\BibitemShut {NoStop}%
\bibitem [{\citenamefont {von Neumann}(1931)}]{neumann_1931}%
  \BibitemOpen
  \bibfield  {author} {\bibinfo {author} {\bibfnamefont {J.}~\bibnamefont {von
  Neumann}},\ }\href {\doibase 10.1007/BF01457956} {\bibfield  {journal}
  {\bibinfo  {journal} {Math. Annal.}\ }\textbf {\bibinfo {volume} {104}},\
  \bibinfo {pages} {570} (\bibinfo {year} {1931})}\BibitemShut {NoStop}%
\bibitem [{\citenamefont {von Neumann}(1996)}]{neumann_book}%
  \BibitemOpen
  \bibfield  {author} {\bibinfo {author} {\bibfnamefont {J.}~\bibnamefont {von
  Neumann}},\ }\href@noop {} {\emph {\bibinfo {title} {Mathematische Grundlagen
  der Quantenmechanik}}},\ \bibinfo {edition} {2nd}\ ed.\ (\bibinfo
  {publisher} {Springer},\ \bibinfo {year} {1996})\BibitemShut {NoStop}%
\bibitem [{\citenamefont {Balian}(1981)}]{balian_low_theorem_balian_1981}%
  \BibitemOpen
  \bibfield  {author} {\bibinfo {author} {\bibfnamefont {R.}~\bibnamefont
  {Balian}},\ }\href@noop {} {\bibfield  {journal} {\bibinfo  {journal} {C. R.
  Acad. Sci. Paris}\ }\textbf {\bibinfo {volume} {292}},\ \bibinfo {pages}
  {1357} (\bibinfo {year} {1981})}\BibitemShut {NoStop}%
\bibitem [{\citenamefont {Low}(1985)}]{balian_low_theorem_low_1985}%
  \BibitemOpen
  \bibfield  {author} {\bibinfo {author} {\bibfnamefont {F.}~\bibnamefont
  {Low}},\ }\enquote {\bibinfo {title} {A passion for physics: Essays in honor
  of {Geoffrey Chew}},}\ \ (\bibinfo  {publisher} {World Scientific},\ \bibinfo
  {year} {1985})\ Chap.\ \bibinfo {chapter} {Complete sets of wave packets},
  pp.\ \bibinfo {pages} {17--22}\BibitemShut {NoStop}%
\bibitem [{\citenamefont
  {Battle}(1988)}]{balian_low_theorem_proof_battle_1988}%
  \BibitemOpen
  \bibfield  {author} {\bibinfo {author} {\bibfnamefont {G.}~\bibnamefont
  {Battle}},\ }\href {\doibase 10.1007/bf00397840} {\bibfield  {journal}
  {\bibinfo  {journal} {Lett. Math. Phys.}\ }\textbf {\bibinfo {volume} {15}},\
  \bibinfo {pages} {175–177} (\bibinfo {year} {1988})}\BibitemShut {NoStop}%
\bibitem [{\citenamefont {Marston}\ and\ \citenamefont
  {Balint-Kurti}(1989)}]{fgh_marston_balint-kurti_1989}%
  \BibitemOpen
  \bibfield  {author} {\bibinfo {author} {\bibfnamefont {C.~C.}\ \bibnamefont
  {Marston}}\ and\ \bibinfo {author} {\bibfnamefont {G.~G.}\ \bibnamefont
  {Balint-Kurti}},\ }\href {\doibase 10.1063/1.456888} {\bibfield  {journal}
  {\bibinfo  {journal} {J. Chem. Phys.}\ }\textbf {\bibinfo {volume} {91}},\
  \bibinfo {pages} {3571} (\bibinfo {year} {1989})}\BibitemShut {NoStop}%
\bibitem [{\citenamefont {Daubechies}\ and\ \citenamefont
  {Janssen}(1993)}]{balian_low_generalisation_janssen_1993}%
  \BibitemOpen
  \bibfield  {author} {\bibinfo {author} {\bibfnamefont {I.}~\bibnamefont
  {Daubechies}}\ and\ \bibinfo {author} {\bibfnamefont {A.~J. E.~M.}\
  \bibnamefont {Janssen}},\ }\href {\doibase 10.1109/18.179336} {\bibfield
  {journal} {\bibinfo  {journal} {IEEE Trans. Inf. Theory}\ }\textbf {\bibinfo
  {volume} {39}},\ \bibinfo {pages} {3} (\bibinfo {year} {1993})}\BibitemShut
  {NoStop}%
\bibitem [{\citenamefont {Machnes}\ \emph {et~al.}(2016)\citenamefont
  {Machnes}, \citenamefont {Assémat}, \citenamefont {Larsson},\ and\
  \citenamefont {Tannor}}]{pvb_math_tannor_2016}%
  \BibitemOpen
  \bibfield  {author} {\bibinfo {author} {\bibfnamefont {S.}~\bibnamefont
  {Machnes}}, \bibinfo {author} {\bibfnamefont {E.}~\bibnamefont {Assémat}},
  \bibinfo {author} {\bibfnamefont {H.~R.}\ \bibnamefont {Larsson}}, \ and\
  \bibinfo {author} {\bibfnamefont {D.~J.}\ \bibnamefont {Tannor}},\ }\href
  {\doibase 10.1021/acs.jpca.5b12370} {\bibfield  {journal} {\bibinfo
  {journal} {J. Phys. Chem. A}\ }\textbf {\bibinfo {volume} {120}},\ \bibinfo
  {pages} {3296} (\bibinfo {year} {2016})}\BibitemShut {NoStop}%
\bibitem [{\citenamefont {Manthe}\ and\ \citenamefont
  {Köppel}(1990)}]{DVR_matvec_calc_manthe_1990}%
  \BibitemOpen
  \bibfield  {author} {\bibinfo {author} {\bibfnamefont {U.}~\bibnamefont
  {Manthe}}\ and\ \bibinfo {author} {\bibfnamefont {H.}~\bibnamefont
  {Köppel}},\ }\href {\doibase 10.1063/1.459606} {\bibfield  {journal}
  {\bibinfo  {journal} {J. Chem. Phys.}\ }\textbf {\bibinfo {volume} {93}},\
  \bibinfo {pages} {345} (\bibinfo {year} {1990})}\BibitemShut {NoStop}%
\bibitem [{\citenamefont {Bramley}\ and\ \citenamefont
  {Carrington}(1993)}]{DVR_matvec_calc_carrington_1993}%
  \BibitemOpen
  \bibfield  {author} {\bibinfo {author} {\bibfnamefont {M.~J.}\ \bibnamefont
  {Bramley}}\ and\ \bibinfo {author} {\bibfnamefont {T.}~\bibnamefont
  {Carrington}},\ }\href {\doibase 10.1063/1.465576} {\bibfield  {journal}
  {\bibinfo  {journal} {J. Chem. Phys.}\ }\textbf {\bibinfo {volume} {99}},\
  \bibinfo {pages} {8519} (\bibinfo {year} {1993})}\BibitemShut {NoStop}%
\bibitem [{\citenamefont {Wang}\ and\ \citenamefont
  {Carrington~Jr.}(2001)}]{efficient_matvec_pruned_carrington_2001}%
  \BibitemOpen
  \bibfield  {author} {\bibinfo {author} {\bibfnamefont {X.-G.}\ \bibnamefont
  {Wang}}\ and\ \bibinfo {author} {\bibfnamefont {T.}~\bibnamefont
  {Carrington~Jr.}},\ }\href {\doibase 10.1021/jp003792s} {\bibfield  {journal}
  {\bibinfo  {journal} {J. Phys. Chem. A}\ }\textbf {\bibinfo {volume} {105}},\
  \bibinfo {pages} {2575} (\bibinfo {year} {2001})}\BibitemShut {NoStop}%
\bibitem [{\citenamefont {Brown}\ and\ \citenamefont
  {Carrington}(2015)}]{pvb_H2O_carrington_2015}%
  \BibitemOpen
  \bibfield  {author} {\bibinfo {author} {\bibfnamefont {J.}~\bibnamefont
  {Brown}}\ and\ \bibinfo {author} {\bibfnamefont {T.}~\bibnamefont
  {Carrington}},\ }\href {\doibase 10.1063/1.4926805} {\bibfield  {journal}
  {\bibinfo  {journal} {J. Chem. Phys.}\ }\textbf {\bibinfo {volume} {143}},\
  \bibinfo {pages} {044104} (\bibinfo {year} {2015})}\BibitemShut {NoStop}%
\bibitem [{\citenamefont {Wilson}(1987)}]{gabor_basis_wilson_1987}%
  \BibitemOpen
  \bibfield  {author} {\bibinfo {author} {\bibfnamefont {K.~G.}\ \bibnamefont
  {Wilson}},\ }\href@noop {} {\enquote {\bibinfo {title} {Generalized wannier
  functions},}\ }\bibinfo {howpublished} {Cornell University preprint}
  (\bibinfo {year} {1987})\BibitemShut {NoStop}%
\bibitem [{\citenamefont {Daubechies}, \citenamefont {Jaffard},\ and\
  \citenamefont {Journé}(1991)}]{wilson_gabor_basis_journe_1991}%
  \BibitemOpen
  \bibfield  {author} {\bibinfo {author} {\bibfnamefont {I.}~\bibnamefont
  {Daubechies}}, \bibinfo {author} {\bibfnamefont {S.}~\bibnamefont {Jaffard}},
  \ and\ \bibinfo {author} {\bibfnamefont {J.-L.}\ \bibnamefont {Journé}},\
  }\href {\doibase 10.1137/0522035} {\bibfield  {journal} {\bibinfo  {journal}
  {SIAM J. Math. Anal.}\ }\textbf {\bibinfo {volume} {22}},\ \bibinfo {pages}
  {554} (\bibinfo {year} {1991})}\BibitemShut {NoStop}%
\bibitem [{\citenamefont
  {Löwdin}(1950)}]{lowdin_orthogonalization_lowdin_1950}%
  \BibitemOpen
  \bibfield  {author} {\bibinfo {author} {\bibfnamefont {P.}~\bibnamefont
  {Löwdin}},\ }\href {\doibase http://dx.doi.org/10.1063/1.1747632} {\bibfield
   {journal} {\bibinfo  {journal} {J. Chem. Phys.}\ }\textbf {\bibinfo {volume}
  {18}},\ \bibinfo {pages} {365} (\bibinfo {year} {1950})}\BibitemShut
  {NoStop}%
\bibitem [{\citenamefont {Machnes}, \citenamefont {Assémat},\ and\
  \citenamefont {Tannor}(2016)}]{pvb_algorithms_tannor_2016}%
  \BibitemOpen
  \bibfield  {author} {\bibinfo {author} {\bibfnamefont {S.}~\bibnamefont
  {Machnes}}, \bibinfo {author} {\bibfnamefont {E.}~\bibnamefont {Assémat}}, \
  and\ \bibinfo {author} {\bibfnamefont {D.}~\bibnamefont {Tannor}},\
  }\href@noop {} {\enquote {\bibinfo {title} {{Quantum Dynamics in Phase space
  using the Biorthogonal von Neumann bases: Algorithmic Considerations}},}\ }
  (\bibinfo {year} {2016}),\ \Eprint {http://arxiv.org/abs/arXiv:1603.03963}
  {arXiv:1603.03963} \BibitemShut {NoStop}%
\bibitem [{Note1()}]{Note1}%
  \BibitemOpen
  \bibinfo {note} {We use \protect \texttt {std::unordered\protect \_set} of
  the C++ programming language standard library as implemented in the GNU
  compiler collection\cite {gcc_manual}}\BibitemShut {NoStop}%
\bibitem [{\citenamefont {Sedgewick}\ and\ \citenamefont
  {Wayne}(2011)}]{sedgewick_book}%
  \BibitemOpen
  \bibfield  {author} {\bibinfo {author} {\bibfnamefont {R.}~\bibnamefont
  {Sedgewick}}\ and\ \bibinfo {author} {\bibfnamefont {K.}~\bibnamefont
  {Wayne}},\ }\href@noop {} {\emph {\bibinfo {title} {Algorithms}}},\ \bibinfo
  {edition} {4th}\ ed.\ (\bibinfo  {publisher} {Addison-Wesley Professional},\
  \bibinfo {year} {2011})\BibitemShut {NoStop}%
\bibitem [{\citenamefont {Park}\ and\ \citenamefont
  {Light}(1986)}]{sil_light_1986}%
  \BibitemOpen
  \bibfield  {author} {\bibinfo {author} {\bibfnamefont {T.~J.}\ \bibnamefont
  {Park}}\ and\ \bibinfo {author} {\bibfnamefont {J.~C.}\ \bibnamefont
  {Light}},\ }\href {\doibase 10.1063/1.451548} {\bibfield  {journal} {\bibinfo
   {journal} {J. Chem. Phys.}\ }\textbf {\bibinfo {volume} {85}},\ \bibinfo
  {pages} {5870} (\bibinfo {year} {1986})}\BibitemShut {NoStop}%
\bibitem [{\citenamefont {Worth}\ \emph {et~al.}()\citenamefont {Worth},
  \citenamefont {Beck}, \citenamefont {J{\"a}ckle},\ and\ \citenamefont
  {Meyer}}]{mctdh_package}%
  \BibitemOpen
  \bibfield  {author} {\bibinfo {author} {\bibfnamefont {G.~A.}\ \bibnamefont
  {Worth}}, \bibinfo {author} {\bibfnamefont {M.~H.}\ \bibnamefont {Beck}},
  \bibinfo {author} {\bibfnamefont {A.}~\bibnamefont {J{\"a}ckle}}, \ and\
  \bibinfo {author} {\bibfnamefont {H.-D.}\ \bibnamefont {Meyer}},\ }\href@noop
  {} {}\bibinfo {howpublished} {The {MCTDH} {P}ackage, {V}ersion 8.4.10 (2014).
  {S}ee http://mctdh.uni-hd.de}\BibitemShut {NoStop}%
\bibitem [{\citenamefont {Engel}(1992)}]{acorr_2t_engel_1992}%
  \BibitemOpen
  \bibfield  {author} {\bibinfo {author} {\bibfnamefont {V.}~\bibnamefont
  {Engel}},\ }\href {\doibase http://dx.doi.org/10.1016/0009-2614(92)85155-4}
  {\bibfield  {journal} {\bibinfo  {journal} {Chem. Phys. Lett.}\ }\textbf
  {\bibinfo {volume} {189}},\ \bibinfo {pages} {76 } (\bibinfo {year}
  {1992})}\BibitemShut {NoStop}%
\bibitem [{\citenamefont {Manthe}, \citenamefont {Meyer},\ and\ \citenamefont
  {Cederbaum}(1992)}]{NO2_mctdh_cederbaum_1992}%
  \BibitemOpen
  \bibfield  {author} {\bibinfo {author} {\bibfnamefont {U.}~\bibnamefont
  {Manthe}}, \bibinfo {author} {\bibfnamefont {H.-D.}\ \bibnamefont {Meyer}}, \
  and\ \bibinfo {author} {\bibfnamefont {L.~S.}\ \bibnamefont {Cederbaum}},\
  }\href {\doibase 10.1063/1.463332} {\bibfield  {journal} {\bibinfo  {journal}
  {J. Chem. Phys.}\ }\textbf {\bibinfo {volume} {97}},\ \bibinfo {pages} {9062}
  (\bibinfo {year} {1992})}\BibitemShut {NoStop}%
\bibitem [{\citenamefont {Worth}, \citenamefont {Meyer},\ and\ \citenamefont
  {Cederbaum}(1998)}]{pyrazine_24d_cederbaum_1998}%
  \BibitemOpen
  \bibfield  {author} {\bibinfo {author} {\bibfnamefont {G.~A.}\ \bibnamefont
  {Worth}}, \bibinfo {author} {\bibfnamefont {H.-D.}\ \bibnamefont {Meyer}}, \
  and\ \bibinfo {author} {\bibfnamefont {L.~S.}\ \bibnamefont {Cederbaum}},\
  }\href {\doibase 10.1063/1.476947} {\bibfield  {journal} {\bibinfo  {journal}
  {J. Chem. Phys.}\ }\textbf {\bibinfo {volume} {109}},\ \bibinfo {pages}
  {3518} (\bibinfo {year} {1998})}\BibitemShut {NoStop}%
\bibitem [{\citenamefont {Shimshovitz}\ and\ \citenamefont
  {Tannor}(2012{\natexlab{b}})}]{pvb_wavelet_tannor_2012}%
  \BibitemOpen
  \bibfield  {author} {\bibinfo {author} {\bibfnamefont {A.}~\bibnamefont
  {Shimshovitz}}\ and\ \bibinfo {author} {\bibfnamefont {D.~J.}\ \bibnamefont
  {Tannor}},\ }\href {\doibase 10.1063/1.4751484} {\bibfield  {journal}
  {\bibinfo  {journal} {J. Chem. Phys.}\ }\textbf {\bibinfo {volume} {137}},\
  \bibinfo {pages} {101103} (\bibinfo {year} {2012}{\natexlab{b}})}\BibitemShut
  {NoStop}%
\bibitem [{\citenamefont {Westermann}\ and\ \citenamefont
  {Manthe}(2012)}]{correlated_vN_entropy_effective_Hamiltonian_manthe_2012}%
  \BibitemOpen
  \bibfield  {author} {\bibinfo {author} {\bibfnamefont {T.}~\bibnamefont
  {Westermann}}\ and\ \bibinfo {author} {\bibfnamefont {U.}~\bibnamefont
  {Manthe}},\ }\href {\doibase 10.1063/1.4720567} {\bibfield  {journal}
  {\bibinfo  {journal} {J. Chem. Phys.}\ }\textbf {\bibinfo {volume} {136}},\
  \bibinfo {pages} {204116} (\bibinfo {year} {2012})}\BibitemShut {NoStop}%
\bibitem [{\citenamefont {Hirsch}, \citenamefont {Buenker},\ and\ \citenamefont
  {Petrongolo}(1991)}]{no2_pes_petrongolo_1991}%
  \BibitemOpen
  \bibfield  {author} {\bibinfo {author} {\bibfnamefont {G.}~\bibnamefont
  {Hirsch}}, \bibinfo {author} {\bibfnamefont {R.~J.}\ \bibnamefont {Buenker}},
  \ and\ \bibinfo {author} {\bibfnamefont {C.}~\bibnamefont {Petrongolo}},\
  }\href {\doibase 10.1080/00268979100101791} {\bibfield  {journal} {\bibinfo
  {journal} {Mol. Phys.}\ }\textbf {\bibinfo {volume} {73}},\ \bibinfo {pages}
  {1085–1099} (\bibinfo {year} {1991})}\BibitemShut {NoStop}%
\bibitem [{\citenamefont {Leonardi}\ \emph {et~al.}(1994)\citenamefont
  {Leonardi}, \citenamefont {Petrongolo}, \citenamefont {Keshari},
  \citenamefont {Hirsch},\ and\ \citenamefont
  {Buenker}}]{no2_pes_buenker_1994}%
  \BibitemOpen
  \bibfield  {author} {\bibinfo {author} {\bibfnamefont {E.}~\bibnamefont
  {Leonardi}}, \bibinfo {author} {\bibfnamefont {C.}~\bibnamefont
  {Petrongolo}}, \bibinfo {author} {\bibfnamefont {V.}~\bibnamefont {Keshari}},
  \bibinfo {author} {\bibfnamefont {G.}~\bibnamefont {Hirsch}}, \ and\ \bibinfo
  {author} {\bibfnamefont {R.~J.}\ \bibnamefont {Buenker}},\ }\href {\doibase
  10.1080/00268979400100414} {\bibfield  {journal} {\bibinfo  {journal} {Mol.
  Phys.}\ }\textbf {\bibinfo {volume} {82}},\ \bibinfo {pages} {553–565}
  (\bibinfo {year} {1994})}\BibitemShut {NoStop}%
\bibitem [{Note2()}]{Note2}%
  \BibitemOpen
  \bibinfo {note} {Brown and Carrington noted problems with the condition
  number of $\protect \textbf {G}$, namely numbers $\sim 10^3$ after an
  adaption of the width of each von Neumann function for a grid of size
  $48$.\cite {pvb_H2O_carrington_2015} We do not experience such
  problems.}\BibitemShut {Stop}%
\bibitem [{\citenamefont {Raab}\ \emph {et~al.}(1999)\citenamefont {Raab},
  \citenamefont {Worth}, \citenamefont {Meyer},\ and\ \citenamefont
  {Cederbaum}}]{pyrazine_24d_cederbaum_1999}%
  \BibitemOpen
  \bibfield  {author} {\bibinfo {author} {\bibfnamefont {A.}~\bibnamefont
  {Raab}}, \bibinfo {author} {\bibfnamefont {G.~A.}\ \bibnamefont {Worth}},
  \bibinfo {author} {\bibfnamefont {H.-D.}\ \bibnamefont {Meyer}}, \ and\
  \bibinfo {author} {\bibfnamefont {L.~S.}\ \bibnamefont {Cederbaum}},\ }\href
  {\doibase 10.1063/1.478061} {\bibfield  {journal} {\bibinfo  {journal} {J.
  Chem. Phys.}\ }\textbf {\bibinfo {volume} {110}},\ \bibinfo {pages} {936}
  (\bibinfo {year} {1999})}\BibitemShut {NoStop}%
\bibitem [{\citenamefont {Zanchet}, \citenamefont {Roncero},\ and\
  \citenamefont {Bulut}(2016)}]{S+_H2_scattering_bulut_2016}%
  \BibitemOpen
  \bibfield  {author} {\bibinfo {author} {\bibfnamefont {A.}~\bibnamefont
  {Zanchet}}, \bibinfo {author} {\bibfnamefont {O.}~\bibnamefont {Roncero}}, \
  and\ \bibinfo {author} {\bibfnamefont {N.}~\bibnamefont {Bulut}},\ }\href
  {\doibase 10.1039/C6CP00604C} {\bibfield  {journal} {\bibinfo  {journal}
  {Phys. Chem. Chem. Phys.}\ }\textbf {\bibinfo {volume} {18}},\ \bibinfo
  {pages} {11391} (\bibinfo {year} {2016})}\BibitemShut {NoStop}%
\bibitem [{\citenamefont {Ma}, \citenamefont {Li},\ and\ \citenamefont
  {Guo}(2012)}]{HO_CO_dyn_guo_2012}%
  \BibitemOpen
  \bibfield  {author} {\bibinfo {author} {\bibfnamefont {J.}~\bibnamefont
  {Ma}}, \bibinfo {author} {\bibfnamefont {J.}~\bibnamefont {Li}}, \ and\
  \bibinfo {author} {\bibfnamefont {H.}~\bibnamefont {Guo}},\ }\href {\doibase
  10.1021/jz301064w} {\bibfield  {journal} {\bibinfo  {journal} {J. Phys. Chem.
  Lett.}\ }\textbf {\bibinfo {volume} {3}},\ \bibinfo {pages} {2482} (\bibinfo
  {year} {2012})}\BibitemShut {NoStop}%
\bibitem [{\citenamefont
  {Worth}(2000)}]{mctdh_selected_configurations_worth_2000}%
  \BibitemOpen
  \bibfield  {author} {\bibinfo {author} {\bibfnamefont {G.~A.}\ \bibnamefont
  {Worth}},\ }\href {\doibase http://dx.doi.org/10.1063/1.481438} {\bibfield
  {journal} {\bibinfo  {journal} {J. Chem. Phys.}\ }\textbf {\bibinfo {volume}
  {112}},\ \bibinfo {pages} {8322} (\bibinfo {year} {2000})}\BibitemShut
  {NoStop}%
\bibitem [{\citenamefont {Napoli}\ \emph {et~al.}(2014)\citenamefont {Napoli},
  \citenamefont {Fabregat-Traver}, \citenamefont {Quintana-Ortí},\ and\
  \citenamefont {Bientinesi}}]{tensor_contractions_BLAS_usage_bientinesi_2014}%
  \BibitemOpen
  \bibfield  {author} {\bibinfo {author} {\bibfnamefont {E.~D.}\ \bibnamefont
  {Napoli}}, \bibinfo {author} {\bibfnamefont {D.}~\bibnamefont
  {Fabregat-Traver}}, \bibinfo {author} {\bibfnamefont {G.}~\bibnamefont
  {Quintana-Ortí}}, \ and\ \bibinfo {author} {\bibfnamefont {P.}~\bibnamefont
  {Bientinesi}},\ }\href {\doibase http://dx.doi.org/10.1016/j.amc.2014.02.051}
  {\bibfield  {journal} {\bibinfo  {journal} {Appl. Math. Comput.}\ }\textbf
  {\bibinfo {volume} {235}},\ \bibinfo {pages} {454 } (\bibinfo {year}
  {2014})}\BibitemShut {NoStop}%
\bibitem [{\citenamefont {Goto}\ and\ \citenamefont
  {Geijn}(2008)}]{gemm_goto_2008}%
  \BibitemOpen
  \bibfield  {author} {\bibinfo {author} {\bibfnamefont {K.}~\bibnamefont
  {Goto}}\ and\ \bibinfo {author} {\bibfnamefont {R.~A. v.~d.}\ \bibnamefont
  {Geijn}},\ }\href {\doibase 10.1145/1356052.1356053} {\bibfield  {journal}
  {\bibinfo  {journal} {ACM Trans. Math. Softw.}\ }\textbf {\bibinfo {volume}
  {34}},\ \bibinfo {pages} {12:1} (\bibinfo {year} {2008})}\BibitemShut
  {NoStop}%
\bibitem [{\citenamefont {Cooper}\ and\ \citenamefont
  {Carrington}(2009)}]{pruned_prod_basis_mapping_carrington_2009}%
  \BibitemOpen
  \bibfield  {author} {\bibinfo {author} {\bibfnamefont {J.}~\bibnamefont
  {Cooper}}\ and\ \bibinfo {author} {\bibfnamefont {T.}~\bibnamefont
  {Carrington}},\ }\href {\doibase 10.1063/1.3140272} {\bibfield  {journal}
  {\bibinfo  {journal} {J. Chem. Phys.}\ }\textbf {\bibinfo {volume} {130}},\
  \bibinfo {pages} {214110} (\bibinfo {year} {2009})}\BibitemShut {NoStop}%
\bibitem [{\citenamefont {Stallman}\ \emph {et~al.}(2015)\citenamefont
  {Stallman} \emph {et~al.}}]{gcc_manual}%
  \BibitemOpen
  \bibfield  {author} {\bibinfo {author} {\bibfnamefont {R.~M.}\ \bibnamefont
  {Stallman}} \emph {et~al.},\ }\href {https://gcc.gnu.org/onlinedocs/5.3.0/}
  {\emph {\bibinfo {title} {Using The Gnu Compiler Collection: A Gnu Manual For
  GCC Version 5.3.0}}}\ (\bibinfo {year} {2015})\BibitemShut {NoStop}%
\end{thebibliography}
\end{document}